\documentstyle[12pt,psfig]{article}

\makeatletter

\@addtoreset{equation}{section}
\makeatother
\def\IN{\relax{\rm I\kern-.18em N}}
\def\IR{\relax{\rm I\kern-.18em R}}

\font\cmss=cmss12 \font\cmsss=cmss12 at 7pt
\def\IZ{\relax\ifmmode\mathchoice
{\hbox{\cmss Z\kern-.4em Z}}{\hbox{\cmss Z\kern-.4em Z}}
{\lower.9pt\hbox{\cmsss Z\kern-.4em Z}}
{\lower1.2pt\hbox{\cmsss Z\kern-.4em Z}}\else{\cmss Z\kern-.4em Z}\fi}

\def\inbar{\,\vrule height1.5ex width.4pt depth0pt}
\def\IC{\relax\hbox{$\inbar\kern-.3em{\rm C}$}}

\newcommand{\Od}{{\cal O}}

\newcommand{\pabar}{\not{\!{\partial}}}

\newcommand{\curr}{j_\mu (x)}

\newcommand{\prodj}{\prod_{j=1}^{2n}}

\newcommand{\prodjint}{\prodj\int_0^\beta d\tau_j \int_0^L dq_j}

\newcommand{\sumn}{\sum_{n=0}^{\infty}}

\newcommand{\be}{\begin{equation}}
\newcommand{\ee}{\end{equation}}
\newcommand{\ba}{\begin{eqnarray}}
\newcommand{\ea}{\end{eqnarray}}

\newcommand{\dbibitem}[1]{\bibitem{#1}}
\newcommand{\gsim}{\raise.3ex\hbox{$>$\kern-.75em\lower1ex\hbox{$\sim$}}}
\newcommand{\lsim}{\raise.3ex\hbox{$<$\kern-.75em\lower1ex\hbox{$\sim$}}}

\newcommand{\paa}{\partial}

\newcommand{\NP}[1]{{\it Nucl.\ Phys.\ }{\bf #1}}

\newcommand{\PL}[1]{{\em Phys.\ Lett.\ }{\bf #1}}

\newcommand{\CMP}[1]{{\em Comm.\ Math.\ Phys.\ }{\bf #1}}

\newcommand{\PR}[1]{{\em Phys.\ Rev.\ }{\bf #1}}

\begin{document}

\typeout{--- Title page start ---}

\renewcommand{\thefootnote}{\fnsymbol{footnote}}

\begin{flushright}
IMPERIAL/TP/98-99/51 \\
DAMTP-1999-63 \\
%\today \\
%\\ (LaTeX-ed on \today )
\end{flushright}
\vskip 12pt

\begin{center}

{\large\bf Chiral symmetry restoration in the massive Thirring model
at \\
\vskip 4pt
finite $T$ and $\mu$: Dimensional reduction and the Coulomb gas}
\vskip 1.2cm
{\large
 A.G\'{o}mez Nicola$^a$\footnote{E-mail:
{\tt gomez@eucmax.sim.ucm.es}},
R.J.Rivers$^{b}$\footnote{E-mail: {\tt
R.Rivers@ic.ac.uk}} and D.A.Steer$^{c}$\footnote{E-mail: {\tt
D.A.Steer@damtp.cam.ac.uk}}  }\\
\vskip 5pt
\vskip 3pt
{\it a})
Departamento de F\'{\i}sica
Te\'orica\\ Universidad Complutense, 28040, Madrid, Spain \\
\vskip 3pt
{\it b})
Theoretical Physics,
Blackett Laboratory, Imperial College, Prince Consort Road,
\\ London, SW7 2BZ, U.K.\\
\vskip 3pt
{\it c}) D.A.M.T.P., Silver Street, Cambridge, CB3 9EW, U.K.\\
\end{center}

\vskip 1.2cm

%See pp.175
\renewcommand{\thefootnote}{\arabic{footnote}}
\setcounter{footnote}{0}
\typeout{--- Main Text Start ---}

\begin{abstract}

We show that in certain limits the (1+1)-dimensional massive Thirring
model at finite temperature $T$ is equivalent to a one-dimensional
Coulomb gas of charged particles at the same $T$.  This equivalence is
then used to explore the phase structure of the massive Thirring
model.  For strong coupling and $T$ $\gg m$ (the fermion mass), the
system is shown to behave as a free gas of ``molecules'' (charge pairs 
in the Coulomb gas terminology) made of pairs
of chiral condensates.  This binding of chiral condensates is responsible 
for the restoration of chiral symmetry as $T\rightarrow\infty$.
In addition, when a fermion chemical
potential $\mu\neq 0$ is included, the analogy
with a Coulomb gas still holds with $\mu$ playing the r\^{o}le of a
purely imaginary external electric field. For small $T$ and $\mu$  we find
a typical massive Fermi gas behaviour for the fermion density,
whereas for large $\mu$ it shows chiral restoration by means of a  vanishing
effective fermion  mass. Some similarities with the chiral properties
of low-energy QCD at finite $T$ and baryon chemical potential are
discussed.

\end{abstract}

\vspace{1cm}

\section{Introduction}\label{sec:intro}

The massive Thirring (MT) model in two space-time dimensions has
been widely studied as a toy counterpart to low-energy QCD, since it 
does not include many of the complications  arising in 3+1
dimensions.  Amongst the features shared by the MT model and QCD is
hadronisation (bosonisation).  In this primitive version, the MT model 
is equivalent to the sine-Gordon (SG) model, both at zero and 
non-zero temperatures \cite{Coleman,Belg,gs98}. 
Viewed as a non-linear sigma model (NLSM) in 1+1 dimension for a 
single Goldstone boson with an explicit symmetry breaking term, the SG 
model mimics the chiral Lagrangian for low-energy, strongly coupled
QCD (whose r\^ole is played here by the MT model)
 where the lowest energy excitations of the vacuum are the
Goldstone bosons. In addition, the solitonic
 excitations of the SG model (kinks, for brevity)
correspond to Thirring fundamental
fermions and hence they are the analogue of QCD baryons or,
alternatively, the NLSM  skyrmions \cite{sk61}. 

This paper is the second in a series concerning the statistical
mechanical features of those models. In \cite{gs98}, the MT/SG system 
and their equivalences (bosonisation) were analysed at 
finite temperature $T>0$ and non-zero fermion number chemical 
potential $\mu\neq 0$. 
It was shown that when there is a
non-zero net fermion number 
in the
MT model, a topological term arises in the dual SG model which counts
the number of kinks minus antikinks \cite{gs98}.   Physically this 
term represents the thermal bosonisation of fermions into kinks and 
it reflects the existence of pure fermion excitations in the thermal
bath.  The same term appears in the exactly solvable massless Thirring
model \cite{AlvGom98}, and a similar contribution for a nonzero baryon 
chemical potential in the low-energy QCD chiral Lagrangian was
obtained in \cite{AlvGom95}.  However, the massless Thirring model, dual
to the Schwinger model, has a much less rich and relevant structure if 
we have QCD in mind, and we shall not consider it here, except as a 
limiting case. 

 The purpose of the present paper is to analyse the
thermodynamics of the MT model, calculating physical quantities such
as the pressure, the  fermion condensate (the order parameter
of the chiral symmetry) and the fermion number density.  We believe this
to be important for obtaining a better understanding of crucial
physical phenomena such as the QCD chiral phase transition, not only
at finite temperature but also at finite baryon density \cite{bielefeld}.

We now introduce the model.  In Euclidean space-time with
metric $(+,+)$ the Lagrangian density of the MT model
is
$$
{\cal{L}}_{MT} [\bar\psi,\psi] =  -\bar{\psi} (\pabar + m_0)\psi +
\frac{1}{2}g^2  j_\mu (x) j^\mu (x)  .
%\label{MT}
$$
Here  $\psi$ is a two-component
fermionic field, the $0$ subscript denotes bare quantities and
$\curr=\bar\psi (x)\gamma_{\mu} \psi (x)$ where the Euclidean $\gamma$ 
matrices may be found in \cite{gs98}.   
For $g^{2}>0$ forces are attractive and the theory is 
super-renormalisable.  For $-1/2<g^{2}/\pi\leq 0$, further 
renormalisations need to be carried out as discussed
in reference \cite{amit} for example, whereas for $g^2/\pi\leq -1/2$ the
theory is no longer renormalisable \cite{zj}. Throughout this paper we
take $g^{2}>0$ unless otherwise stated.  With QCD in mind we are interested in
strong coupling.  In this we are aided by the general result that
the natural parameter in which to express results is $(1 + g^{2}/\pi
)^{-1}$. 
This is understood
by the explicit nature of the duality with the SG model, 
with Lagrangian density
$$
{\cal L}_{SG}[\phi] = \frac{1}{2} \paa_\mu \phi  \paa^\mu \phi -
\frac{\alpha_0}{\lambda^2} \cos \lambda \phi 
$$
where $\phi$ is a real scalar field.
Recall that these
models are equivalent only in the weak  sense, i.e.\ for vacuum
expectation values and thermal averages, provided that their
renormalised constants are related through \cite{Coleman}
\ba
\frac{\lambda^2}{4\pi}&=&\frac{1}{1+g^2/\pi},
\label{equivconst1}
\\
\frac{\alpha}{\lambda^2}&=&\rho \, m ,
\label{equivconst2}
\ea
where $\rho$ is the renormalisation scale and $m$ the renormalised
mass at that scale. Thus large positive $g^{2}$ corresponds to small
$\lambda^{2}$, for which approximations can be well controlled.

With QCD in mind, we will be interested in the chiral properties of the
MT model as well as of  physical quantities such as the pressure. 
The chiral $U(1)$ transformations are
$\psi\rightarrow \exp(ia\gamma^5)\psi$ for the fermions and
$\phi\rightarrow \phi+a/\lambda$ for the bosons with $a$ real
arbitrary. The massless Thirring model is chiral invariant but the
fermion mass term breaks that symmetry explicitly. Similarly the free
boson theory is chiral invariant, whereas the $\cos\lambda\phi$
term in the SG
model breaks it. Both models have still a residual $\IZ$ invariance,
corresponding to the choice $a=2\pi n$ with $n$ any integer. It is
important to remark that in two space-time dimensions, the breaking of
a continuous symmetry ($U(1)\rightarrow \IZ$ in this case)
cannot be spontaneous \cite{col73}. The $U(1)$ and $\IZ$ symmetries
are,  respectively, the
counterparts of the  chiral and isospin symmetries for QCD, $\alpha_0$ and
$\lambda$ playing the r\^ole of the pion mass squared and the inverse
of the pion decay constant respectively in the effective chiral
Lagrangian to lowest order, i.e.\ the NLSM
\cite{Angelbook}.
These generalities apart, the identification (\ref{equivconst1})
which, for strong coupling becomes $\lambda g\approx 2\pi$, is the
only attribute of the SG
theory that we shall use. Depending on which is most convenient, we
shall switch between the use of $g$ and $\lambda$. 
This is not to say that the SG model has no interest in its own
right. It fact, it is an ideal testbed for summation schemes for the
pressure and, as such, will be considered elsewhere \cite{adr2}.  It is 
also of direct relevance to the study of kinks in Josephson junctions \cite{Roberto}.

In order to make progress we comment on a further property of the 2D MT
model that also has a counterpart in QCD: at $T=0$ and
$\mu=0$ it is also equivalent to a 2D classical
statistical-mechanical system which consists of non-relativistic
particles of charge $\pm q$ (a Coulomb gas) 
at a temperature $T_{CG}$ \cite{amit,Stu}. 
Before clarifying the form of this analogy,
we comment that one of the aims of this paper is to understand
whether a similar equivalence holds when the MT model is
heated  to a non-zero temperature $T$ and for $\mu \neq 0$.  If
it does, we will then be able to use the simpler Coulomb gas
system to explore its phase space.
It has not yet proved possible to perform similar calculations for 
QCD, as represented in \cite{jai}, in which the Coulomb gas is
composed of monopoles.

Indeed, one of  our conclusions 
is that at high temperature $T$, the MT model 
can be equivalent to a {\em one}
dimensional Coulomb gas of particles at the {\em same} temperature
$T$.  This is not the case for the $T=0$ MT model, for which 

$$
T_{CG} = \frac{2 \pi q^2}{\lambda^2}
$$
leading to a Kosterlitz-Thouless (metal-insulator) 
transition at $\lambda^2 = 8 \pi$ \cite{Stu}.
The 1D Coulomb gas model has been solved exactly \cite{lenard1,lenard2}
and we use those results to extract information on the behaviour of observables
such as the pressure and chiral condensate of the MT model.
Furthermore, when $\mu\neq 0$ in the MT model, we show that
the analogy with a 1D Coulomb gas still holds with $\mu$ playing the 
r\^{o}le of a purely imaginary external electric field.

This paper is organised as follows. In section
\ref{sec:dimred} we analyse the conditions under which the equivalence
which a 1D Coulomb gas holds.  Having specified this ``dimensionally reduced''
regime, the exact link between the parameters
of the different models is then specified in section \ref{sec:pf} where
the pressure of the MT model is also calculated.  Section \ref{sec:sigmas}
is concerned with the fermion condensate.  We see that for
large $g$ and
$T$ $\gg m$, the MT model
behaves as a free gas of ``molecules'' made of pairs of chiral
condensates. This binding of chiral condensates is responsible 
for the restoration of chiral symmetry as $T\rightarrow\infty$,
and no phase transition takes place.  The effects of a fermion chemical
potential $\mu\neq 0$ on the pressure and fermion density are
 discussed in section \ref{sec:chempot}.

\section{Dimensional reduction of the $T>0$  MT
model.}\label{sec:dimred}

At nonzero temperature $T$ and zero chemical potential $\mu = 0$ the
 MT model partition function is given by \cite{Belg,gs98} 
\be
Z_{MT} (T)=Z_0^F(T)\sumn \left(\frac{1}{n!}\right)^2
\left[\frac{\rho \, m}{2}\left(\frac{T}
{\rho}\right)^{\lambda^2/4\pi}\right]^{2n} F_{2n}(\lambda,T,L).
\label{sgparfun}
\ee
Here $Z_0^F(T)$ is the partition function for a
two-dimensional free massless Fermi gas, which, ignoring irrelevant
vacuum terms, is given by
$$
Z_0^F(T)= \exp \left[\frac{\pi L
T}{6}\right],
$$
and the function $F_{2n}(\lambda,T,L)$ is
\be
 F_{2n}(\lambda,T,L) = \prodjint
\prod_{k<j}\left[ Q^2(x_j-x_k)\right]^{\epsilon_j\epsilon_k
\lambda^2/4\pi}.
\label{Fdef}
\ee
Here $x_j\equiv(\tau_j,q_j)$, $\beta=T^{-1}$,
$$
\epsilon_j=\left\{ \begin{array}{cl} +&j=1,\dots,n\\
                                     -&j=n+1,\dots,2n
\end{array}\right. 
$$
and  $L$ is the length of the
system. We will eventually take the thermodynamic or
$L\rightarrow\infty$ limit in all our results. It must be clarified
though that the $L\rightarrow\infty$
limit will be taken always keeping $\beta$
finite. In fact, we will see explicit examples below in which the
$L\rightarrow\infty$ and $T\rightarrow 0^+$ limit do not commute.
%%%
%
Finally, the $Q$ variable in (\ref{Fdef}) also appears in the finite
temperature free massless boson and fermion propagators \cite{Belg} and
is given by
\be
Q^2(q,\tau) 
= \sinh(\frac{\pi (q +i\tau )}{\beta})\sinh(\frac{\pi (q- i\tau )}{\beta})
\label{qs}
\ee
betraying its conformal origins.
As discussed in \cite{gs98}, the integrals in (\ref{sgparfun}) are
convergent for $g^{2}>0$.

Equation (\ref{sgparfun})  will be our starting point here. First,
notice that
\be
 Q^2(q,\tau)\longrightarrow \frac{1}{4}\exp{\frac{2\pi \vert q\vert}{\beta}}
  \qquad \mbox{for} \quad \frac{\pi \vert q\vert}{\beta}\gg 1 \qquad
\forall \tau.
\label{qasym}
\ee
The key observation is therefore that if we were allowed to replace the
$Q^2$  functions in (\ref{sgparfun}) by their asymptotic values
(\ref{qasym}) then, with appropriate definitions, (\ref{sgparfun}) 
would resemble the grand
canonical partition function of a one-dimensional classical gas of
charged particles with positions labelled by $q_{i}$.  Remember that the
two-particle Coulomb potential for charges $\pm \sigma$ at points
$q_1$ and $q_2$ on the line is  $V \propto \pm \sigma^2 |q_1 - q_2|$.
This 1D Coulomb gas system was studied a long time ago
\cite{lenard1,lenard2} and it can be solved exactly.  We shall explore
this analogy later on in section \ref{sec:pf} and use the exact results to
calculate the pressure and chiral condensate of the MT model.

Before doing so, however, we need to establish the
conditions under which the  replacement given in (\ref{qasym})---which
we will denote dimensional reduction (DR)---can be
safely made. This DR regime is the 2D analogue of
more complicated situations analysed in the thermal field theory
literature, which roughly works for high temperatures and large
distances \cite{gi80,appi81,ka96}.

 Clearly the regions in the integrand
 (\ref{sgparfun}) where the replacement (\ref{qasym}) is {\em not}
allowed are those where $|q_j-q_k|\leq \beta/\pi$. In those regions
the dominant contributions to the integral come from $x_j\simeq x_k$
with $\epsilon_j\epsilon_k=-1$ since
$$
 Q^2(q,\tau)\longrightarrow (\pi T)^2 (q^2+\tau^2)
  \qquad \mbox{for} \quad (q,\tau)\rightarrow (0^+,0^+).
$$
Thus, on the one hand, we expect
 that for high enough $T$ the contribution (arising in the denominator)
can be neglected.
 On the other hand, the above contributions become
 more important as $g^{2}$ decreases, so that one may also think that
 for large enough $g^{2}$ the approach could be equally justified.
 Bearing these considerations in mind, we shall analyse
 the limits of high $T$ and large $g^{2}$ separately in sections
 \ref{ssec:hight} and \ref{ssec:smallla}.

Before proceeding, two comments are in order regarding the
 expression (\ref{sgparfun}).  First, every  term in the $n$-sum
 picks up an overall $\beta L$ factor. In other words, the number of
 two-dimensional  independent variables in the integral is actually
$2n-1$. This is most easily seen by changing variables to
\be
z_1=x_1-x_2,z_2=x_2-x_3,\dots,z_{2n-1}=x_{2n-1}-x_{2n},z_{2n}=x_{2n}
\label{covxz}
\ee
so that in (\ref{sgparfun})
$$
x_j-x_k=\sum_{i=j}^{k-1} z_i
$$
and therefore the integrand is independent of $z_{2n}$, which yields the
 $\beta L$ factor. Notice that this is typical of
 closed loops  in
 perturbation theory \cite{kapusta}. Order by order
 one has to consider all possible connected closed diagrams
for the partition function and the
 $\beta L$ factor is just the consequence of total energy-momentum
 conservation. One should bear in mind that
 the physically relevant object
 is not the partition function but the pressure, defined in the
 thermodynamic limit as
\be
P=\lim_{L\rightarrow\infty} \frac{1}{\beta L}\log Z_{MT} (L,T)
\label{press}
\ee
which behaves as an intensive quantity.

The second comment concerns the scale dependence.
The partition function (\ref{sgparfun}) is scale independent (and so
is the pressure) since the
 explicit dependence on the renormalisation scale $\rho$ is exactly
cancelled by the implicit
 dependence of the mass $m (\rho)$ (see \cite{gs98}
 for details). Thus, unless otherwise stated and whenever dealing with 
scale-independent objects in the following, we
will choose for convenience $\rho=m$, the renormalised mass of the Thirring
 fermion.

\subsection{High $T$ limit}\label{ssec:hight}

\begin{figure}
\hspace*{-1.5cm}
\vspace*{-4cm}
\psfig{figure=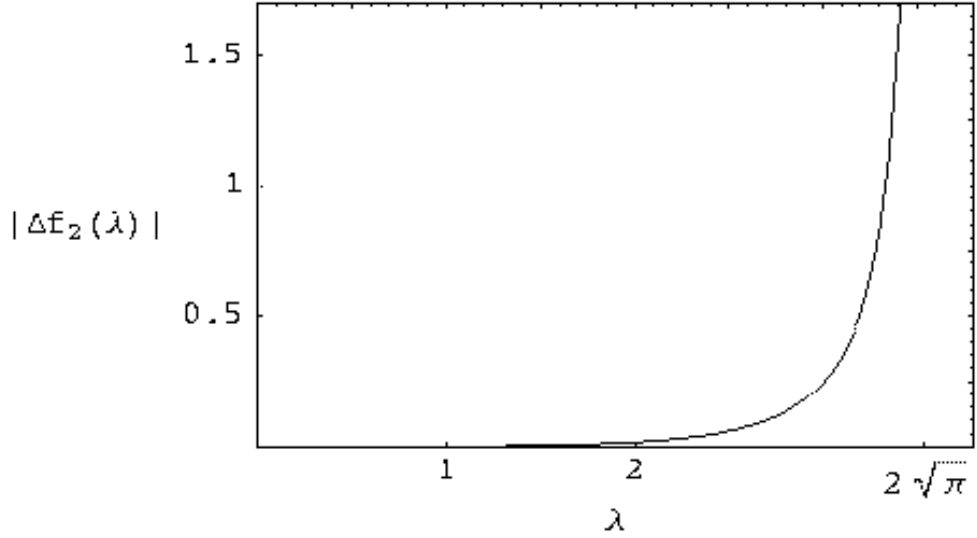,height=22cm}
\vspace*{-9cm}
\caption {\label{fig1} The function  $\Delta f_2 (\lambda)$ }
\end{figure}

Let us rescale $q_j\rightarrow q_j\beta$ and $\tau_j\rightarrow
\tau_j\beta$ $\forall j$ in (\ref{Fdef})\footnote{For much of 
this section it is more convenient to work
with $\lambda$ rather than $g$ in the first instance.}.
 Then all the relevant $T$ dependence is now outside the integrals
in (\ref{Fdef}) since in the thermodynamic limit
$\eta=L/\beta\rightarrow
 \infty$ the integrals are finite as long as $\lambda^2<4\pi$---excluding
  of course the $\eta$ overall factor mentioned above.  Then, choosing
 $\rho=m$, the series in  (\ref{sgparfun}) yields effectively a perturbative
 series in $(m^2/T^2)^{(\pi + 
2g^{2})/(2\pi + 2g^{2})}$ = $(m^2/T^2)^{1-\lambda^2/8\pi}$:
$$
Z_{MT}(T)=Z_0^F(T)\left\{1+\sum_{n=1}^{\infty} \left(\frac{1}{n!}\right)^2
\left[2\left(\frac{m^2}{T^2}\right)^{1-\lambda^2/8\pi}\right]^{2n}
 \frac{1}{4}\eta f_{2n} (\lambda)\right\}
$$
with $f_{2n} (\lambda)$ independent of $\eta$ in the thermodynamic
limit and given by
$$
 f_{2n}(\lambda) = \prod_{j=1}^{2n-1}\int_0^1 dz_j^0\int_0^\infty
 dz_j^1 
\prod_{k<j+1}
\left[ Q^2(\beta \sum_{i=k}^j z_i)\right]^{\epsilon_{j+1}\epsilon_k
\lambda^2/4\pi},
$$
where $z_j\equiv (z_j^0,z_j^1)$ is defined  in  (\ref{covxz}). 
It follows from (\ref{qs}) that 
$Q^2(\beta q,\beta\tau)$
is independent of $\beta$. 
Therefore, to leading order in $m/T$, we can write the MT pressure as
\be
P_{MT}(\lambda,T)=P_{MT}^{asym} (\lambda,T)\left[1+
 \frac{6}{\pi}\left(\frac{m}{T}\right)^{4-\lambda^2/2\pi}
\!\!\!\!\!\!
\Delta f_2 (\lambda)\right] +
\Od\left(\frac{m}{T}\right)^{8-\lambda^2/\pi}
\label{pexpla}
\ee
where the $asym$ superscript
 denotes the value obtained by replacing the $Q$'s
 by their asymptotic values in (\ref{qasym}), and
$\Delta f_2 (\lambda)= f_2 (\lambda)- f_2^{asym} (\lambda)$ with
\be
 f_2 (\lambda)=\int_0^\infty dq\int_0^1 d\tau
\left[Q^2(\beta q,\beta\tau)\right]^{-\lambda^2/4\pi}
\label{f2lam}
\ee
and
$$
f_2^{asym} (\lambda)=
\frac{2^{\lambda^2/2\pi+1}}{\lambda^2}.
$$
We have evaluated numerically  $\Delta f_2 (\lambda)$ and the result
is plotted in Figure \ref{fig1}.
 We see that it remains of  $\Od(1)$ until very close  to the
limiting case
 $\lambda^2/4\pi\lsim 1$, or $g^{2}\gsim 0$, where it diverges. Thus,
 the relative error for $T\gg m$
\footnote{Notice that this is a scale dependent condition}
 is at least of order $(m/T)^2$ and we can
therefore conclude that for $T \gg m$, DR
is valid for $g^{2}>0$.

\subsection{Strong coupling (large $g^{2}$) limit.}\label{ssec:smallla}

We would now like to study the pressure for $g^{2}/\pi\gg1$ ($\lambda^2/4\pi\ll 1$). The first
 observation is that the integrals in (\ref{Fdef}) diverge if
we set  $\lambda=0$ and then take the limit $L\rightarrow\infty$. Therefore
 we shall keep $L$  finite and then take it to infinity (keeping
 $\beta$ finite) only when the pressure (\ref{press}) is calculated. 
Then, expanding (\ref{Fdef}) in
$\lambda^2/4\pi$ one obtains
$$
F_{2n}(\lambda,T,L)=(\beta L)^{2n}\left\{1-\frac{n\beta\lambda^2}{\pi L}
 \int_0^\eta dq\int_0^1 d\tau \log\left[Q^2(\beta q,\beta\tau)\right]
 + \Od\left(\frac{\lambda^2}{4\pi}\right)^2\right\}.
$$
Thus
$$
F_{2n}(\lambda,T,L)=F_{2n}^{asym}(\lambda,T,L)-
n(\beta L)^{2n}\frac{\lambda^2}{4\pi}\hat I(\eta)+
\Od\left(\frac{\lambda^2}{4\pi}\right)^2
$$
with
\be
\hat I(\eta)=\frac{4}{\eta}
\left[\int_0^\eta dq\int_0^1 d\tau \log\left[Q^2(\beta q,\beta\tau)\right]
+2\eta\log 2-\pi\eta^2  \right]
\label{ieta}
\ee
so that
\ba
Z_{MT} (T,\lambda,L)&=&Z_{MT}^{asym} (T,\lambda,L)
-\frac{\lambda^2}{4\pi}\hat I(\eta)Z_0^F(T)
\sumn n\left(\frac{1}{n!}\right)^2
\left[\frac{m^2}{2}\beta L\right]^{2n}
+ \Od\left(\frac{\lambda^2}{4\pi}\right)^2
\nonumber\\
&=&Z_{MT}^{asym} (T,\lambda,L)-
\frac{\lambda^2}{4\pi}\hat I(\eta)Z_0^F(T)\frac{z}{2}I_0'(z)
+ \Od\left(\frac{\lambda^2}{4\pi}\right)^2
\nonumber
\ea
where
\be
z=m^2\beta L\left(\frac{T}{m}\right)^{\lambda^2/4\pi}
\label{zz}
\ee
and $I_0 (z)$ is the modified Bessel function of
 zeroth order. On the other hand,
\be
Z_{MT} (T,\lambda,L)=Z_0^F(T)\sumn \left(\frac{1}{n!}\right)^2
\left(\frac{z}{2}\right)^{2n} +\Od\left(\frac{\lambda^2}{4\pi}\right)=
Z_0^F(T)I_0(z)+\Od\left(\frac{\lambda^2}{4\pi}\right)
\label{losl}
\ee
and $Z_{MT}^{asym} (T,\lambda,L)$ has the same leading order in
$\lambda^2/4\pi$ as above.
 Therefore, we can write for the pressure
$$
P_{MT}(T,\lambda)=P_{MT}^{asym} (T,\lambda)-\lim_{L\rightarrow\infty}
\frac{1}{\beta L}\left[\frac{\lambda^2}{8\pi}\hat I(\eta)z
\frac{I_0'(z)}{I_0(z)}\right]
+ \Od\left(\frac{\lambda^2}{4\pi}\right)^2.
$$

Now, for large $z$ \cite{abramo},
$$
I_0(z)=\frac{e^z}{\sqrt{2\pi z}}+\Od\left(\frac{1}{z}\right) \qquad
 \mbox{for}\quad  z\gg 1
$$
so that taking $L\rightarrow\infty$  
\footnote{Note that, in turn, from (\ref{losl}) we find  the
 asymptotic limit $P_{MT}\simeq \frac{\pi}{6}T^2 + m^2$ 
 for small $\lambda$, which we
 will recover in the Coulomb gas approximation in section \ref{sec:pf}.}
we  finally get
\be
P_{MT}(T,\lambda)=P_{MT}^{asym} (T,\lambda)-\frac{m^2\lambda^2}{8\pi}
 \hat I (\eta\rightarrow\infty)+\Od\left(\frac{\lambda^2}
{4\pi}\right)^2.
\label{deltapsg}
\ee
Numerical analysis of the function $\hat{I} (\eta)$ shows  that it
clearly
 vanishes as $\eta\rightarrow\infty$. Therefore we have
 $\Delta P_{MT}=\Od(\lambda^4)$. Although it
 might seem that this result is valid for any $T$,
one has to be extremely careful when approaching
 the limit $T\rightarrow 0^+$. In fact, we see
 that  expanding (\ref{zz}) in $\lambda$ yields logarithmic factors
 $\log T/m$  hidden in the $\Od(\lambda^4)$ in
 (\ref{deltapsg}), but which will show up at next-to-leading order.
Therefore the above results should not be trusted 
for temperatures $T\ll m$. This
 is just a consequence of the non-uniformity of the $T\rightarrow 0^+$
 limit or, in other words, that the $L\rightarrow\infty$ and
 $T\rightarrow 0^+$ do not commute in general.  Hence we will
 consider that our approach is justified  for large $g^2/\pi$
 and $T\gsim m$. Recall that the presence of the $\log T/m$
 factors in the $\lambda$ expansion may be also troublesome  for very
 large $T$, which gives a hint that the high $T$ limit may not be
 compatible with perturbation theory, as has been noticed in more
 complicated situations \cite{rebhan}.
In our case though, we have a well defined
 high $T$ expansion, namely our expansion in
 $(m^2/T^2)^{1-\lambda^2/8\pi}$  discussed in
 section \ref{ssec:hight}, which we
 will indeed be able to resum in the DR regime.

To summarise, the results of this section are that the range of validity of the
 DR approach  which we will be
 considering  for the pressure is
\ba
T\gg m &\quad& , \quad g^{2}>0,\nonumber\\
T\gsim m &\quad& , \quad  \frac{g^{2}}{\pi}\gg 1.
\label{drworks}
\ea
We now turn to the 1D Coulomb gas system and specify exactly the link
between it and the MT model.

\section{The MT partition function as a Coulomb gas}\label{sec:pf}

Consider the partition function for a one-dimensional neutral classical
non-relativistic gas of charged particles. In particular, take $N$
positively charged particles and $N$  negative ones, where the
magnitude of the charge will be denoted by $\sigma$.  In one dimension, 
these charges interact via a Coulomb potential proportional to
$\pm\sigma^2$ and to the distance between them on the line. This system can be
also interpreted as uniformly charged plane sheets moving along the
direction normal to their planes \cite{lenard1,lenard2}.
Then, the partition function in the grand canonical ensemble
 at fixed temperature $\theta$, fugacity $z$ and length $L$ is
 \cite{lenard1,lenard2}
\be
\Omega (z,\theta,\sigma,L)=\sum_{N=0}^{\infty}
\frac{ z^{2N}}{(N!)^2}
\left(\prod_{i=1}^{2N}\int_0^L dq_i\right)
\exp\left[2\pi\sigma^2\theta^{-1}\sum_{1\leq j<i\leq 2N}
\epsilon_i\epsilon_j \vert q_i-q_j\vert\right]
\label{coulparfun}
\ee
where $\epsilon_i=1$ for $i\leq N$ and $\epsilon_i=-1$ for $i>N$. The
fugacity is related to the chemical potential $\mu$ by 
$z=\sqrt{2\pi M\theta}\exp\theta^{-1}\mu$, where $M$ is the mass of
the particles\footnote{This definition of
fugacity differs from the usual one by the factor of $\sqrt{M}$.  We
have absorbed this factor into the definition here as $M$ plays no r\^ole in 
future considerations.}.
However, we will keep $z$ instead of $\mu$
here, so as not to confuse this chemical potential with the fermion
chemical potential we will introduce in section \ref{sec:chempot}.
In the thermodynamic limit the pressure and the mean particle density 
are given by
\ba
P(z,\theta,\sigma)&=&\lim_{L\rightarrow\infty}
\frac{\theta}{L}\log \Omega (z,\theta,\sigma,L)
\nonumber
\\
n(z,\theta,\sigma)&=&\frac{\ll 2N \gg}{L}=
\theta^{-1} z\frac{\partial}{\partial z}
P(z,\theta,\sigma)
\nonumber
\ea
where $\ll \cdot \gg$ denotes the statistical average in the above ensemble.
We therefore realise that by making the replacement (\ref{qasym}) in
 (\ref{sgparfun}) and setting $\rho=m$, one can write\footnote{Again 
we work first with $\lambda$ and translate to $g$ later.}
\be
Z_{MT}(T,L)=Z_0^F(T,L)
\Omega\left(z=\frac{m^2}{2T}\left(\frac{2T}{m}\right)^{\lambda^2/4\pi},
\theta=T,\sigma=\frac{\lambda T}{\sqrt{4\pi}},L\right).
\label{pfequiv}
\ee

This is the central equivalence of this paper. Its utility lies in the
fact that the above classical problem was solved analytically in
references \cite{lenard1,lenard2}.  There it was shown that for small
$\theta$ the 1D Coulomb gas behaves  as a gas of free
``molecules'', made  up of $+-$ charges pairs bound together (since 
the mean  kinetic energy is much smaller than the mean 
potential energy). The
main features of this phase are: i) The pressure is small compared to
$2\pi\sigma^2$, which represents the pressure between $+-$ charges, and 
ii) The probability density of finding a $+-$ pair within a given
distance $r$ is much bigger than that of finding a $++$ pair for
distances $r$ less than a typical ``molecule'' size, which on the
other hand is much smaller than the typical inter-particle distance. On
the contrary, for large $\theta$ (when the mean  kinetic
energy is much larger than the mean  potential energy)
 the charges are completely
deconfined, forming an electrically neutral ``plasma'' of $2N$ free
 particles, where the
pressure is higher than in the ``molecule'' phase. The
crucial point  in our case is that, from (\ref{pfequiv}) we see that
 the unit charge {\it grows}
with temperature ($\sigma\propto T$) and therefore the above phases are
going to be reversed (high $T$ ``molecule'' phase and low $T$
``plasma'' phase), as we shall now see in detail.

Let us first  quote  the result for the
pressure in \cite{lenard2}\footnote{Note that in \cite{lenard2}, the
pressure and other quantities are calculated in units of
$\theta/2\pi=1$ with $\sigma$ integer. Changing variables in
(\ref{coulparfun}) as $q_i\rightarrow q_i\theta/2\pi\sigma^2$, we have
$\Omega (z,\theta,\sigma,L)=
\Omega(z\theta/2\pi\sigma^2,2\pi,1,2\pi\sigma^2\theta^{-1}L)$ so that
Edwards and 
Lenard's results can easily be rescaled for our purposes.}
$$
P(z,\theta,\sigma)=2\pi\sigma^2
\gamma_0\left[\frac{z\theta}{2\pi\sigma^2}\right]
%\label{presscoul}
$$
where $\gamma_0(\hat z)$ is the highest eigenvalue of  Mathieu's
 differential equation 
\be
\left[\frac{d^2}{d\phi^2}+2\hat z \cos\phi\right]y(\phi)=
\gamma y(\phi)
\label{mathieu}
\ee
with $y(\phi+2\pi)=y(\phi)$. Notice that (\ref{mathieu}) differs
slightly from the notation customarily used for Mathieu's equation
(see for instance \cite{abramo}), which is recovered by  setting
 $\gamma=-a/4$, $\hat{z}=-q/4$ and $\phi=2v$.
Therefore, from (\ref{pfequiv}) we obtain for the full pressure of the
 MT model in the DR regime:
\be
P_{MT}(T)=\frac{\pi}{6}T^2  + P_{C}(T,\lambda)
\label{presstotal}
\ee
with the ``Coulomb gas'' pressure being given by 
$$
P_{C}(T,\lambda)=\frac{\lambda^2 T^2}{2}
\gamma_0\left[\frac{m^2}{\lambda^2 T^2}
\left(\frac{2T}{m}\right)^{\lambda^2/4\pi}\right].
%\label{pressSGcoul}
$$
Or, in terms of $g^{2}$,
\be
P_{C}(T,g)=\frac{2\pi T^2}{1 + g^{2}/\pi}
\gamma_0\left[\frac{m^2}{4\pi T^2}\bigg(1+ \frac{g^{2}}{\pi}\bigg)
\left(\frac{2T}{m}\right)^{(1 + g^{2}/\pi)^{-1}}\right].
\label{pressSGcoulg}
\ee

The eigenvalues and eigenfunctions of the characteristic problem
(\ref{mathieu}) are well known and tabulated. We refer to
\cite{abramo} for their general properties and quote here only
those relevant for our purposes. For instance, the
asymptotic limits of $\gamma_0(\hat z)$ for very large and very small
$\hat z$ are
\ba
\gamma_0 (\hat z)&\simeq& 2\hat z^2 + \Od(\hat z^4) \qquad \mbox{for}
 \quad \hat z\ll 1 \label{mathieuasym1}\\
 \gamma_0 (\hat z)&\simeq& 2\hat z - \sqrt{\hat z} + \Od (1) \qquad \mbox{for}
 \quad \hat z\gg 1. \label{mathieuasym2}
\ea

 The relevant parameter setting the qualitative
behaviour of the system is  the argument of 
$\gamma_0$ in (\ref{pressSGcoulg}),
\be
\hat z
=\frac{m^2}{4\pi T^2}\bigg(1+ \frac{g^{2}}{\pi}\bigg)
\left(\frac{2T}{m}\right)^{(1 + g^{2}/\pi)^{-1}} \; \; :
\label{hatz}
\ee
for small $\hat z$, the system is in the ``molecule'' phase, whereas
for large $\hat z$ it is in the ``plasma'' phase.
Let us then consider three separate limiting cases:
\begin{itemize}
\item [i)] $g^2\gg \pi$ and $T\gg mg/2\pi$: This is the very
  high $T$ regime and we have $\hat z\ll 1$, so that equations
(\ref{mathieuasym1}) and (\ref{pressSGcoulg}) give
  $P_C\sim m^2(m^2 g^2 /4\pi^2T^2)$,
i.e.\ the Coulomb pressure vanishes at high
  $T$ for large $g^2$. This is the ``molecule'' phase described in
\cite{lenard1} in which the Coulomb charges tend to pair,
thus lowering the pressure. As noted above,  this
``unexpected'' behaviour of high temperature binding is just a
consequence of the fact that our coupling ``charge'' $\sigma$ in
(\ref{pfequiv}) {\em increases} with $T$. Besides, when adding
the free term in (\ref{presstotal}), the total pressure is an
increasing function of
$T$, behaving in the high $T$ (and large $g^2$) regime as a free
gas of massless fermions. This is indeed the first indication that the
``molecule'' phase corresponds in fact to a phase of asymptotic chiral
symmetry restoration. The reason is that we expect that if the chiral
symmetry is restored, the system behaves roughly as the massless case
which is chiral invariant. In other words, we would expect the
effective fermion mass to vanish in the $T\rightarrow \infty$ limit
when the system tends to restore the chiral symmetry.

\item [ii)] $g^2\gg \pi$ but $T\ll mg/2\pi$: Now
equation (\ref{mathieuasym2}) gives $P_C\sim m^2-\pi m T/g$. Since
$\hat z\gg 1$, now we are in the ``plasma'' phase. 
 Notice that this same result, to leading order,
 could have been obtained  through our analysis for large $g$  in
 section \ref{ssec:smallla}, c.f.\ equation (\ref{losl}). Again we
 note that to $\Od(\lambda^2)$ in the expansion of $P_C$ there is a
 contribution proportional to $\lambda^2 m^2 \log (T/m)$ which
 prevents us from going to $T\ll m$.  This is
 another reason why in the small $\lambda$ (large $g$)
  regime we can only trust DR for $T\gsim m$, which is the region
covered in this second case. 

\item  [iii)] $0<g^2 \ll 1$
and $2\pi T\gg m$: we have now
$\hat z\simeq   m/2\pi T$.
 Again, $\hat z\ll 1$ and we are in the ``molecular'' phase.
Remember that in this region we can only trust
  DR  for $T\gg m$. Thus
  $P_C\sim m^2/\pi$ so that the pressure tends to a constant value
  for high $T$. Actually, if we were able to push our results formally 
    to $g^{2}\leq 0$  $(\lambda^2\geq 4\pi)$,  the pressure would 
start growing for large $T$. Clearly, our previous calculations in the 
MT model are
not justified in this case, which at least would require extra
renormalisations. However, from the viewpoint of the Coulomb gas only, 
there is no particular problem with taking $g^2$ negative. In other
words, the dimensionally reduced theory is UV finite. 
Notice also that, as we make $g^{2}$ more negative, the Coulomb
  correction to the free boson gas term becomes more and more
  important. In particular, if  we take  $g^{2}/\pi \lsim -1/2$
\footnote{Recall that this corresponds to $\lambda^2\gsim 8 \pi$ and thus
to the Kosterlitz-Thouless transition in the zero-temperature
theory.},
 $P_C$ would grow
 quadratically with $T$, thus being of the same order as the free
 contribution. 
\end{itemize}

\begin{figure}
\hspace*{-1cm}
\vspace*{-6cm}
\psfig{figure=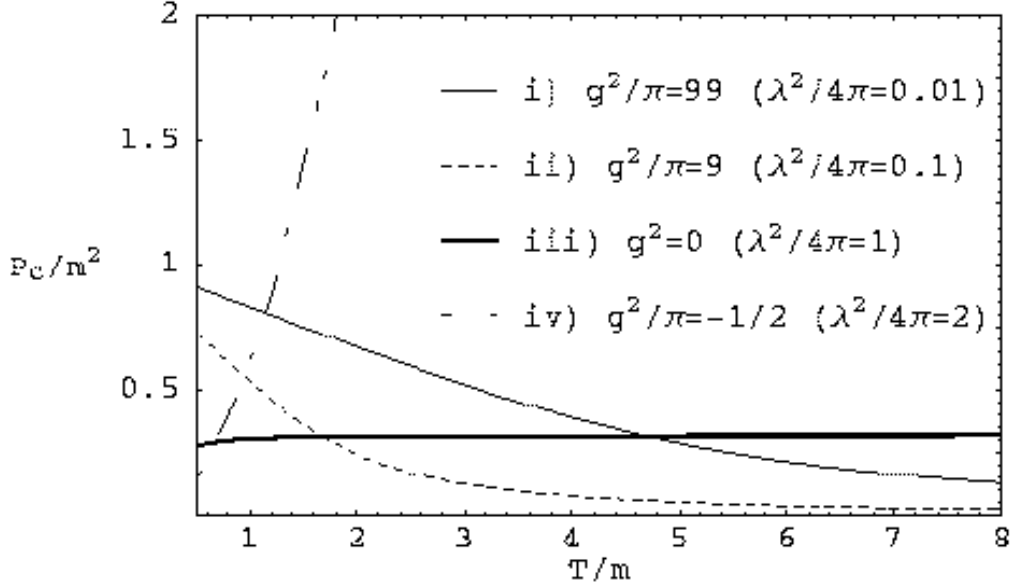,height=22cm}
\vspace*{-6.5cm}
\caption {\label{fig2} The Coulomb pressure $P_C (T,g)$.}
\end{figure}

\begin{figure}
\hspace*{-1cm}
\vspace*{-7.5cm}
\psfig{figure=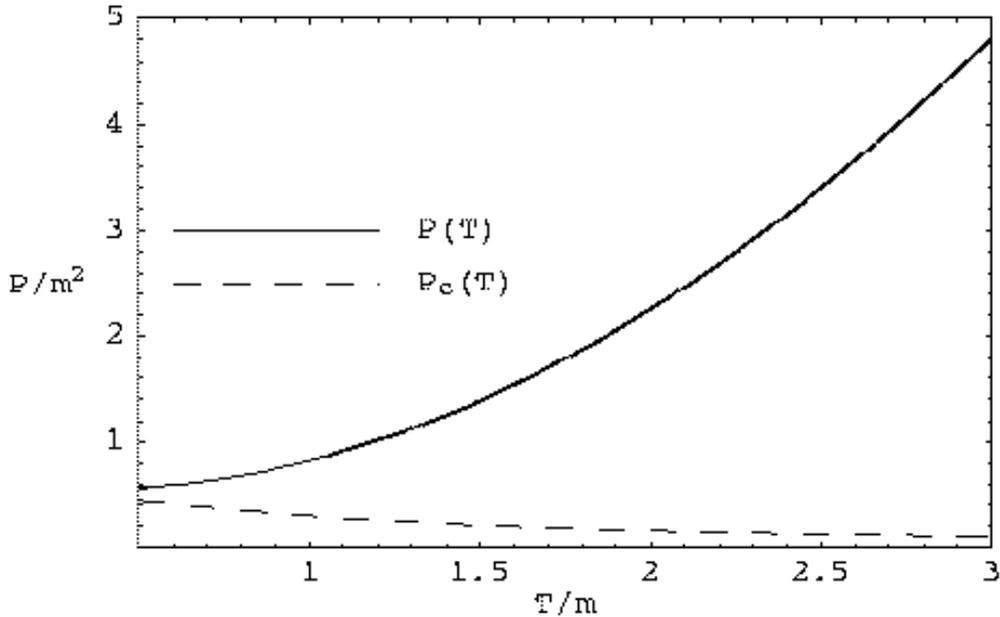,height=22cm}
\vspace*{-5cm}
\caption {\label{fig3} The total pressure and  the Coulomb pressure
  for $g^2/\pi=1$}
\end{figure}

In Figure \ref{fig2} we have plotted $P_C(T)$ in units of $m$
for different values of
$g^{2}$. Notice
 that, as we have said before, we cannot claim that curve iv) corresponds
 to the MT model in the DR regime. We are simply extrapolating the
 Coulomb gas results to that point. As for curve iii), we should intend it
 as the limit of the dimensionally reduced 
 MT model for very small but positive $g^2$,
 where we can still trust our DR approach, as long as we look  only to the 
 $T/m \gg 1$ tail of  that curve.
Note also  that $P_C(T)$ is always a continuous
function of the temperature and thus we do not see a phase transition
in $T$, consistent with being in two dimensions. In Figure \ref{fig3}
 we plot the Coulomb pressure and the total pressure for
 $g^2/\pi=1$ in order to estimate the size of the
 ``Coulomb'' gas correction to the free gas.

\section{The chiral charges and ``molecules''}\label{sec:sigmas}

In the previous section we have exploited the analogy of our
dimensionally reduced MT partition function with that of a Coulomb gas
on the line in order to calculate the pressure. 
In this section we will show that this analogy can be
extended further, i.e.\ that it also works in the same way for other
observables. In particular, we will concentrate on thermal
averages of products of the operators
$$
\sigma_\pm (x)=\frac{1}{2}\overline\psi(1\pm\gamma^5)\psi
$$
since these
account
for the chiral properties of the system.  Under the chiral
transformations discussed in the introduction,
$\sigma_\pm (x)\rightarrow \exp(\pm 2i a)\sigma_\pm (x)$; in other words, the
$\sigma_\pm$ operators have well defined $\pm$ chiral  charge.  Below
we will show that in our
Coulomb gas analogy they play the r\^ole of the charges, and the
chiral invariant combination $\sigma_+\sigma_-$ will then represent a
``molecule''. Thus by forming ``molecules'', the system tends to
restore the chiral symmetry, i.e.\ it behaves as the massless theory
where only $\sigma_+\sigma_-$ combinations are allowed.

We will show below that the $\sigma_\pm$  correlators can be related with the
1D Coulomb gas
reduced density functions discussed in \cite{lenard2}. These functions
are defined as follows: $f_\pm (x) dx$  denotes the probability of
finding a $\pm$ charge in the element $dx$; 
$f_{+-} (x_1,x_2)dx_1 dx_2$ is
the joint probability of finding a $+$ charge in  $dx_1$
and a $-$ one in  $dx_2$, and so on. It turns out that  such density
functions can also be exactly calculated \cite{lenard2}. We will only
consider here the one-point and two-point functions, which will
provide information on the relationship between Coulomb ``molecule''
pairing and the chiral symmetry, as explained above.

\subsection{The fermion condensate and chiral symmetry vs
``molecule'' pairing}\label{ssec:cond}

The first correlator we will analyse is just the one-point function,
i.e.\ the condensate of $\sigma_\pm$. 
Observe that $\ll \sigma_+ \gg \; = \; \ll \sigma_- \gg$
since the MT model is invariant under $\psi_+\leftrightarrow \psi_-$ where
$\psi_\pm=(1/2)(1\pm\gamma^5)\psi$ are the right and left-handed  projections
of the spinor field ($\phi\rightarrow -\phi$ in the SG model). This is a parity
 transformation and has nothing to do with the chiral transformations
 we have been discussing here. We shall implicitly make use of this
 $+\leftrightarrow -$ symmetry in the correlators throughout this
 section. 

The scale invariant fermion condensate $m\ll \overline\psi \psi \gg$
is the order
parameter of the chiral symmetry, exactly as the quark condensate in
QCD.  An immediate
consequence of the absence of a phase transition in 2D
 is that the fermion condensate cannot vanish
strictly at any temperature. On the other hand, for high temperatures
$T\gg m$ we would expect that the mass scale $m$ becomes irrelevant and
thus the system would restore the chiral symmetry, as indeed happens
in the QCD chiral phase transition. This was already suggested by our previous
analysis of the pressure.  We may therefore expect the
condensate to become smaller at large $T$ but never to reach
zero. However, we do not expect  chiral restoration  for
all positive values of $g^2$. An indication that this may actually be
the case is the following.  
In the $g^2\rightarrow 0^+$ limit, the system should
behave as a free massive fermion theory. However,  it is not difficult to see
that for $g=0$ the condensate behaves for $T\gg m$ as
$\ll\bar\psi\psi\gg\simeq (m/\pi)\log (T^2/m^2)$, so that there is
clearly no chiral  restoration in that case. Therefore we
expect that chiral symmetry
restoration is lost   as the value of $g^2$ is decreased. Indeed, we
have seen that behaviour already for the pressure, where the Coulomb
correction to the massless term did not vanish for large $T$ and for very
small $g^2$.

We will now show that in the DR regime the fermion
condensate can be calculated exactly, thanks again to the analogy with the
Coulomb gas. In fact we do not need to appeal to the density
functions in this case since in the thermodynamic limit
$L\rightarrow\infty$ the system is translationally invariant so that
\be
m\ll\overline\psi\psi\gg \; =
m\frac{\partial}{\partial m} P_{MT} (T).
\label{condder}
\ee
Notice in particular that this implies that the regime in which we
are allowed to replace the $Q$'s by their DR values for the condensate
 will be  the same as for the pressure, i.e, (\ref{drworks}).

Thus we only need to differentiate once in the expression obtained for the
MT partition function in the DR regime to get the
condensate. However, in order to clarify the procedure we will follow for
the two point correlator and  to understand better the analogy between
 Coulomb charges and chiral operators,
let us relate the condensate with the one-charge
density functions $f_\pm$ analysed in \cite{lenard2}. Those functions
are given by  $f_+=f_-$ with
 \ba
f_+ (z,\theta,\sigma,L;X)&=&\frac{1}{\Omega}\sum_{N=1}^{\infty}
\frac{ z^{2N}}{(N!)^2}
\left(\prod_{i=1}^{2N}\int_0^L dq_i\right)\sum_{j=1}^{N}\delta(q_j-X)
\nonumber\\
&\times&
\exp\left[2\pi\sigma^2\theta^{-1}\sum_{1\leq j<i\leq 2N}
\epsilon_i\epsilon_j \vert q_i-q_j\vert\right]\nonumber\\
&=&\frac{1}{\Omega}
\sum_{N=1}^{\infty}
\frac{ z^{2N}}{N!(N-1)!}
\left(\prod_{i=1}^{2N}\int_0^L dq_i\right)\delta(q_N-X)
\nonumber\\&\times&
\exp\left[2\pi\sigma^2\theta^{-1}\sum_{1\leq j<i\leq 2N}
\epsilon_i\epsilon_j \vert q_i-q_j\vert\right],
\label{f+}
\ea
where $\Omega$ is the partition function of (\ref{coulparfun}).
We see that in the $L\rightarrow\infty$ limit, a simple shift
$q_i\rightarrow q_i+X$ $\forall i=1,\dots,2N$
ensures that $f_+$ is independent of $X$. In turn, notice that
\be
f_+ (z,\theta,\sigma,L)=\frac{1}{L}\int_0^L dX f_+ (z,\theta,\sigma,L;X)=
\frac{1}{2L}z\frac{\partial}{\partial z} \log \Omega
 (z,\theta,\sigma,L).
\label{f+der}
\ee

Now compare
(\ref{f+}) with the expression obtained in \cite{gs98} for the condensate: 
\ba
m\ll\overline\psi\psi\gg&=&2
\frac{Z_0^F(T)}{Z_{MT}(T)}\sumn \frac{1}{n!(n+1)!}
\left[\frac{\rho \, m}{2}\left(\frac{T}
{\rho}\right)^{\lambda^2/4\pi}\right]^{2(n+1)}
\nonumber\\
&\times&
\prod_{j=1}^{n+1} \int_0^\beta d\tau_j \int_0^L dq_j
\prod_{k<j}\left[ Q^2(x_j-x_k)\right]^{\epsilon_j\epsilon_k\lambda^2/4\pi}.
\nonumber
\ea
Thus, shifting $n\rightarrow n-1$ in the above equation, replacing
the $Q$'s by their asymptotic values (\ref{qasym}), fixing $\rho=m$
 and comparing with
$f_+$ we obtain in the DR regime
\be
m\ll\overline\psi\psi\gg \;=
2Tf_+\left(z=\frac{m^2}{2T}\left(\frac{2T}{m}\right)^{\lambda^2/4\pi},
\theta=T,\sigma=\frac{\lambda T}{\sqrt{4\pi}},L\right).
\label{condvsf+}
\ee

Relation (\ref{condvsf+}) is very interesting indeed since it
 supports the idea  that the r\^ole of the Coulomb charges is played here
by the  $\sigma_\pm$ chiral correlators (in the DR
regime). We will elaborate further on this issue below.

\begin{figure}
\hspace*{-1cm}
\vspace*{-1cm}
\psfig{figure=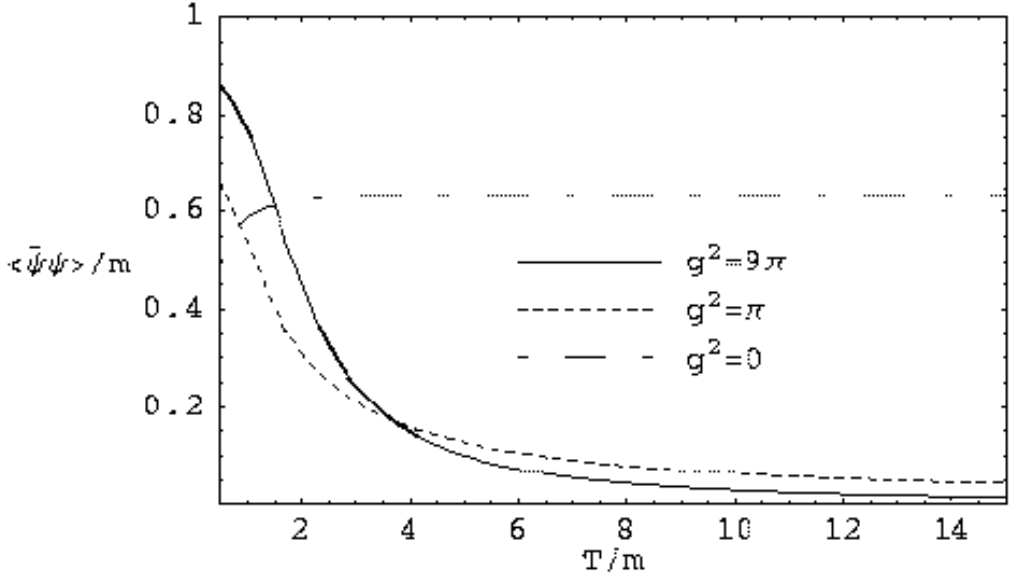,height=22cm}
\vspace*{-11.5cm}
\caption {\label{fig4} The fermion condensate as a function of
  temperature in units of $m$, for different values of $g^2$}
\end{figure}

As it was said above, the fermion condensate can be obtained either
using the analogy with the $f_{\pm}$ functions or by differentiating
 the pressure. We should then be able to rewrite  (\ref{condvsf+}) as 
(\ref{condder}), which would be a consistency check of our
results. For that purpose, all we  need to use is 
 (\ref{f+der}) and the equivalence between $\Omega$ and
$Z_{MT}$ found in the previous section. 
 In doing so it is important to
leave the scale
$\rho$ unfixed, since differentiating with respect to
$m$ is a scale-dependent operation.  After multiplying by $m$ the
result becomes scale independent, and we can safely
fix $\rho=m$. Once that consistency check has been performed, let us
write the final expression for the fermion condensate as 
\be
m\ll\overline\psi\psi\gg=\frac{m^2}{2}
\left(\frac{2T}{m}\right)^{(1 + g^{2}/\pi)^{-1}}
\gamma_0'\left[\frac{m^2}{4\pi T^2}\bigg(1+ \frac{g^{2}}{\pi}\bigg)
\left(\frac{2T}{m}\right)^{(1 + g^{2}/\pi)^{-1}}\right],
\label{cond}
\ee
where $\gamma_0' (z)=d\gamma_0(z)/dz$ and we have converted back to $g^2$.
Remember that unlike the partition function, there is no contribution 
to the condensate coming from the free Bose part 
 since the free theory is chirally symmetric.
Therefore by taking into account the asymptotic behaviour
(\ref{mathieuasym1})-(\ref{mathieuasym2}),  we now analyse
the same limiting cases  considered for the  pressure
in the previous section:

 \begin{itemize}

 \item [i)] $g^2/\pi\gg 1$ and $T\gg mg/2\pi$: we have
$m\ll\overline\psi\psi\gg \; \sim (m^2/2\pi^{2})(mg/T)^2$. The condensate
vanishes asymptotically and hence the chiral
symmetry is restored as $T\rightarrow\infty$.

\item[ii)]  $g^2/\pi\gg 1$ and $T\ll mg/2\pi$: Now 
$m\ll\overline\psi\psi\gg \; \sim m^2 (2T/m)^{(1+g^{2}/\pi)^{-1}}$. This is
the behaviour for temperatures $T\gsim m$.  Notice again that if we
insisted on extrapolating this behaviour to $T=0$ we would find
$\ll\overline\psi\psi\gg (T=0)=0$, which is clearly incorrect since
we know that at $T=0$ the chiral symmetry is broken. Once more, we
remark that we
should trust our result only for $T\gsim m$ due to the presence of
logarithms in the $\lambda$ expansion.

\item[iii)]  $0<g^2 \ll 1$ and $m\ll 2\pi T$:
$m\ll\overline\psi\psi\gg \; \sim 2 m^2/\pi$, so that the condensate tends
to a constant value at high temperatures (just as the pressure did),
and 
there is no chiral symmetry restoration. As we anticipated before, we
do not expect chiral restoration to take place 
for small $g^2$. In turn, notice that 
the exact $g^2=0$ limit predicts 
that the condensate grows logarithmically for large $T$, while we get 
 here that it remains constant for $g^2$ small. This is another sign
 of the special character of the $g^2=0$ case, where, as commented
 several times before, we {\it cannot} apply our DR 
arguments because of the extra UV infinities.

 \end{itemize}
In Figure \ref{fig4}, we have plotted the condensate as a function of
the temperature for different values of $g^{2}$ and we can see the
different  asymptotic limits discussed above. It must be stressed
again that 
the $g^2=0$ case plotted in that figure should be taken as
$g^2$ very small but positive.

From the above discussion
the following picture emerges: for large
values of $g^{2}$, the system at high temperatures behaves like a
gas of neutral ``molecules'' made of pairs of chiral charges
(chiral neutrality) and the chiral
symmetry tends to be restored continuously (i.e, no phase transition)
and asymptotically as $T$ increases.
As we decrease $g^2$, the ``molecule'' phase tends to
disappear in favour of the ``plasma'' phase---as was already 
seen for the pressure.  Here the $\sigma_+$
and $\sigma_-$ correlators can be different from zero in chiral non-invariant
combinations and thus the chiral
symmetry remains broken even for large temperature.

Consequently if we look  now at the  two-point correlators, we
should see a tendency of the system to increase the $\sigma_+\sigma_-$
correlator against the  $\sigma_+\sigma_+$ one in the ``molecule''
phase, i.e, for high temperatures and large enough $g^2$.  That
will be the purpose of the next section.

\subsection{Two-charge correlators and the screening length}\label{ssec:twosig}

Let us begin by recalling the definition of the two-charge density
functions in the Coulomb gas \cite{lenard2}:
\ba
f_{+-}(X,Y)&=&\frac{1}{\Omega}\sum_{N=1}^{\infty}
\frac{ z^{2N}}{(N!)^2}
\left(\prod_{i=1}^{2N}\int_0^L
  dq_i\right)\sum_{i=1}^{N}\sum_{j=N}^{2N}\delta (q_i-X)\delta(q_j-Y)
\nonumber\\
&\times&
\exp\left[2\pi\sigma^2\theta^{-1}\sum_{1\leq j<i\leq 2N}
\epsilon_i\epsilon_j \vert q_i-q_j\vert\right]\nonumber\\
&=&
\frac{1}{\Omega}\sum_{N=1}^{\infty}
\frac{ z^{2N}}{((N-1)!)^2}
\left(\prod_{i=1}^{2N}\int_0^L
  dq_i\right)\delta (q_N-X)\delta(q_{2N}-Y)
\nonumber\\
&\times&
\exp\left[2\pi\sigma^2\theta^{-1}\sum_{1\leq j<i\leq 2N}
\epsilon_i\epsilon_j \vert q_i-q_j\vert\right],
\label{f+-lenard}\\
f_{++}(X,Y)&=&\frac{1}{\Omega}\sum_{N=2}^{\infty}
\frac{ z^{2N}}{(N!)^2}
\left(\prod_{i=1}^{2N}\int_0^L
  dq_i\right)\sum_{i=1}^{N}\sum_{i\neq j=1}^{N}\delta (q_i-X)\delta(q_j-Y)
\nonumber\\
&\times&
\exp\left[2\pi\sigma^2\theta^{-1}\sum_{1\leq j<i\leq 2N}
\epsilon_i\epsilon_j \vert q_i-q_j\vert\right]\nonumber\\
&=&
\frac{1}{\Omega}\sum_{N=2}^{\infty}
\frac{ z^{2N}}{N!(N-2)!}
\left(\prod_{i=1}^{2N}\int_0^L
  dq_i\right)\delta (q_{N-1}-X)\delta(q_{N}-Y)
\nonumber\\
&\times&
\exp\left[2\pi\sigma^2\theta^{-1}\sum_{1\leq j<i\leq 2N}
\epsilon_i\epsilon_j \vert q_i-q_j\vert\right].
\label{f++lenard}
\ea
As in the single charge case,  the above functions  depend only on $X-Y$
in the $L\rightarrow\infty$ limit (translation invariance).
 Also, notice that  $f_{-+}=f_{+-}$ and
$f_{--}=f_{++}$ while for brevity we have omitted the dependences  on
$z,\theta,\sigma,L$. Our task now will be to compare
(\ref{f+-lenard})-(\ref{f++lenard}) with the MT correlators
at finite temperature; 
$\ll\sigma_+\sigma_-\gg$ and $\ll\sigma_+\sigma_+\gg$
respectively. Let us start with the $+-$ correlator. In \cite{gs98},
the following result was obtained:

\ba
\ll T_C \sigma_+ (X)\sigma_- (Y)\gg&=&
\frac{Z_0^F(T)}{Z_{MT}(T)}\left(\frac{T}{\rho}\right)^{\lambda^2/2\pi}
\left(\frac{\rho}{2}\right)^2 \sumn \left(\frac{1}{n!}\right)^2
\left[\frac{\rho \, m}{2}\left(\frac{T}
{\rho}\right)^{\lambda^2/4\pi}\right]^{2n} \nonumber\\
&\times&
\prod_{j=1}^{2(n+1)} \int_0^\beta d\tau_j \int_0^L dq_j
 \delta^{(2)}(x_{n+1}-X)\delta^{(2)}(x_{2n+2}-Y)\nonumber\\
&\times&\prod_{k<j}
\left[ Q^2(x_j-x_k)\right]^{\epsilon_j\epsilon_k\lambda^2/4\pi}
\label{sigma+-sg}
\ea
where $\epsilon_j=+$ for $j=1,\dots,n+1$, $\epsilon_j=-$ for
$j=n+2,\dots,2n+2$ and
$T_C$ means contour ordering along $C=[0,-i\beta]$. 
For convenience we have retained
$\lambda$ in the exponents rather than $g$.  Again, the above correlator
 depends only on $X-Y$ in the thermodynamic limit. Notice that we are
 using the same notation $X$ and $Y$ both for one-dimensional and
 two-dimensional variables; the meaning should become clear from the
 context.

As for the $\sigma_+\sigma_+$ correlator, it can be calculated
through the same procedure followed  in \cite{gs98} for the $+-$
correlators using the
generating functional technique. Notice that this
correlator is not chiral invariant, unlike the   $\sigma_+\sigma_-$
one, which    makes its  analysis slightly more
involved, as explained in \cite{gs98}.
 Despite that, we obtain after
renormalisation
\ba
\ll T_C \sigma_+ (X)\sigma_+ (Y)\gg&=&
\frac{Z_0^F(T)}{Z_{MT}(T)}\left(\frac{T}{\rho}\right)^{\lambda^2/2\pi}
\left(\frac{\rho}{2}\right)^2 \sumn \frac{1}{n!(n+2)!}
\left[\frac{\rho \, m}{2}\left(\frac{T}
{\rho}\right)^{\lambda^2/4\pi}\right]^{2n+2} \nonumber\\
&\times&
\prod_{j=1}^{2(n+2)} \int_0^\beta d\tau_j \int_0^L dq_j
 \delta^{(2)}(x_{n+2}-X)\delta^{(2)}(x_{n+1}-Y)\nonumber\\
&\times&\prod_{k<j}
\left[ Q^2(x_j-x_k)\right]^{\epsilon_j\epsilon_k\lambda^2/4\pi}
\label{sigma++sg}
\ea
where $\epsilon_j=+$ for $j=1,\dots,n+2$, $\epsilon_j=-$ for
$j=n+3,\dots,2n+4$.

Note that the ratio
$$
R(\lambda,T;X)=\frac{\ll T_C \sigma_+ (X)\sigma_- (0)\gg}
{\ll T_C \sigma_+ (X)\sigma_+ (0)\gg}
$$
is scale invariant---this is the observable in which we are
interested. However, before proceeding further, we need to clarify how
 DR works for the above correlators.
We begin by noting that DR   is not allowed
 for all values of $X-Y$ in (\ref{sigma+-sg})-(\ref{sigma++sg}).
In fact, performing the
 $\delta^{(2)}$ integrations, a factor $[Q^2(X-Y)]^{\pm\lambda^2/4\pi}$
 comes outside the integrals and it is clear that the replacement
 (\ref{qasym}) will be allowed for that factor only for
 $q\gg \beta/2\pi$, where $q$ is the spatial component of
 $X-Y$.
 Besides, our analysis for the partition function does not
 imply  that even inside the integrals in
 (\ref{sigma+-sg})-(\ref{sigma++sg})
 such replacement can be done.  The detailed
analysis (which follows a similar line to that of sections 
\ref{ssec:hight} and
\ref{ssec:smallla}) can  be found in the appendix.  
Here we simply summarise the
result:  the structure of the correlators
(\ref{sigma+-sg})-(\ref{sigma++sg}) makes it possible to reduce them
dimensionally (inside the integrals)
in the large $g^{2}$ (small $\lambda$) limit but not in the large $T$
one.
If, in addition one works at large distances $q\gg
\beta/2\pi$, the the $Q$'s outside the
integrals are  also dimensionally reduced.
Bearing this in mind, we shall proceed to evaluate these correlators
in the dimensionally reduced regime by exploiting again the analogy
with the Coulomb gas.

\subsubsection{Two-point correlators and the Coulomb gas}

Given the conditions for DR to work for the
correlators, let us now compare (\ref{sigma+-sg}) with (\ref{f+-lenard}) and
(\ref{sigma++sg}) with  (\ref{f++lenard}). For that purpose we replace
  the $Q$'s inside the integrals in  (\ref{sigma+-sg}) by their
  asymptotic values (\ref{qasym}). In doing so, we have to remember
  that such replacement is not allowed in general for the term
 $j=n+1$, $k=2(n+1)$ (see the above comment).
Next, shift $n\rightarrow n-1$ in the sum
 so that, comparing with (\ref{f+-lenard}),  we  get
\ba
\ll T_C \sigma_+ (X)\sigma_- (0) \gg &=&
\left(\frac{1}{4}\right)^{\lambda^2/4\pi}
\frac{T^2}{m^2}\frac{e^{\lambda^2
    q/2\beta}}{\left[Q^2(q,\tau)\right]^{\lambda^2/4\pi}}\nonumber\\
&\times&
f_{+-}\left(z=\frac{m^2}{2T}\left(\frac{2T}{m}\right)^{\lambda^2/4\pi},
\theta=T,\sigma=\frac{\lambda T}{\sqrt{4\pi}};q\right),
\nonumber
\ea
where $X=(q,\tau)$, we have set $Y=0$ and
we have also made use of (\ref{pfequiv}). It is important to
point out that the overall factor $T^2/m^2$ depends on $\rho$
implicitly through $m$ but does not have explicit $\rho$ dependence
(we {\it cannot} fix $\rho$ for that factor), whereas the
rest---i.e.\ the function $f_{+-}$---is scale independent and we have fixed
$\rho=m$ only in that part.

Following the same steps for the $++$ correlator we now find 
\ba
\ll T_C \sigma_+ (X)\sigma_+ (0) \gg &=&
4^{\lambda^2/4\pi}
\frac{T^2}{m^2}\frac{\left[Q^2(q,\tau)\right]^{\lambda^2/4\pi}}
{e^{\lambda^2    q/2\beta}}\nonumber\\
&\times&
f_{++}\left(z=\frac{m^2}{2T}\left(\frac{2T}{m}\right)^{\lambda^2/4\pi},
\theta=T,\sigma=\frac{\lambda T}{\sqrt{4\pi}};q\right)
%\label{++equiv}
\nonumber
\ea
and therefore the scale-independent ratio of the two yields
\be
R(\lambda,T;X)=2^{-\lambda^2/\pi}\frac{e^{\lambda^2
    q/\beta}}{\left[Q^2(q,\tau)\right]^{\lambda^2/2\pi}}
\frac{f_{+-}\left(z=\frac{m^2}{2T}\left(\frac{2T}{m}\right)^{\lambda^2/4\pi},
\theta=T,\sigma=\frac{\lambda T}{\sqrt{4\pi}};q\right)}
{f_{++}\left(z=\frac{m^2}{2T}\left(\frac{2T}{m}\right)^{\lambda^2/4\pi},
\theta=T,\sigma=\frac{\lambda T}{\sqrt{4\pi}};q\right)}.
\label{r}
\ee

Once more, the main advantage of comparing with the Coulomb gas is
that the functions $f_{++}$ and $f_{+-}$ can be calculated exactly.
They can also be related with the Mathieu equation (\ref{mathieu}) as
 \cite{lenard2}:
$$
f_{+\pm}(z,\theta,\sigma;q)=
\left(\frac{2\pi\sigma^2}{\theta}\right)^2\sum_{m=0}^{\infty} B_m^+
B_m^{\pm} e^{-\left[\gamma_0 (\hat z)-\gamma_m (\hat z)\right]\hat q}.
%\label{lenresfs}
%\ee
$$
Here
$$
 B_m^{\pm}=\hat z\int_{-\pi}^{\pi} d\phi \; y_0 (\phi) y_m (\phi)
e^{\pm i\phi},
$$
the $L\rightarrow\infty$ limit has been taken, $\hat z=\theta
z/2\pi\sigma^2$, $\hat q=2\pi\sigma^2 q/\theta$ and $y_m (\phi)$, $\gamma_m$
are, respectively,  the eigenfunctions and eigenvalues of
(\ref{mathieu}), with $\gamma_{m-1}>\gamma_m$, $m=1,2,\dots$.

First, note that since   $\gamma_m<0$ for $m\geq 1$ and $\gamma_0>0$,
only the $B_0$ term survives in the
above sum in the limit $q\rightarrow\infty$. Then, since
$y_0(\phi)=y_0(-\phi)$, we have
$f_{+\pm}\rightarrow (B_0)^2$ in that limit. In fact $B_0=\hat z\gamma_0'(\hat
z)$. Since in that limit we can also dimensionally reduce the $Q^2$
factors in (\ref{r}), we get
%\be
$$
R(\lambda,T;q\rightarrow\infty,\tau)=1.
%\label{asR}
%\ee
$$

\begin{figure}
\hspace*{-2cm}
\hbox{\psfig{figure=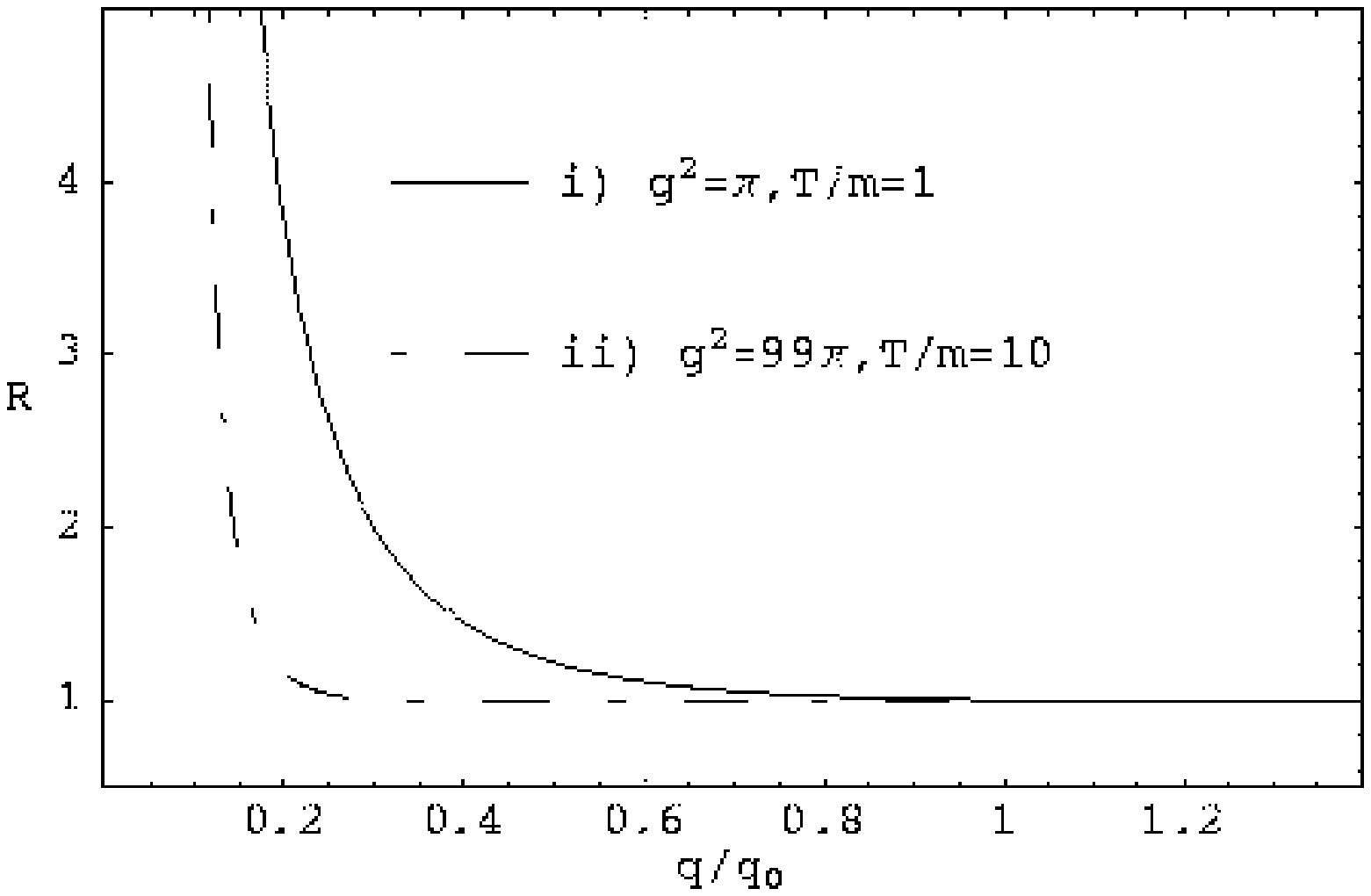,width=10cm,height=15cm}\hspace*{-1cm}
\psfig{figure=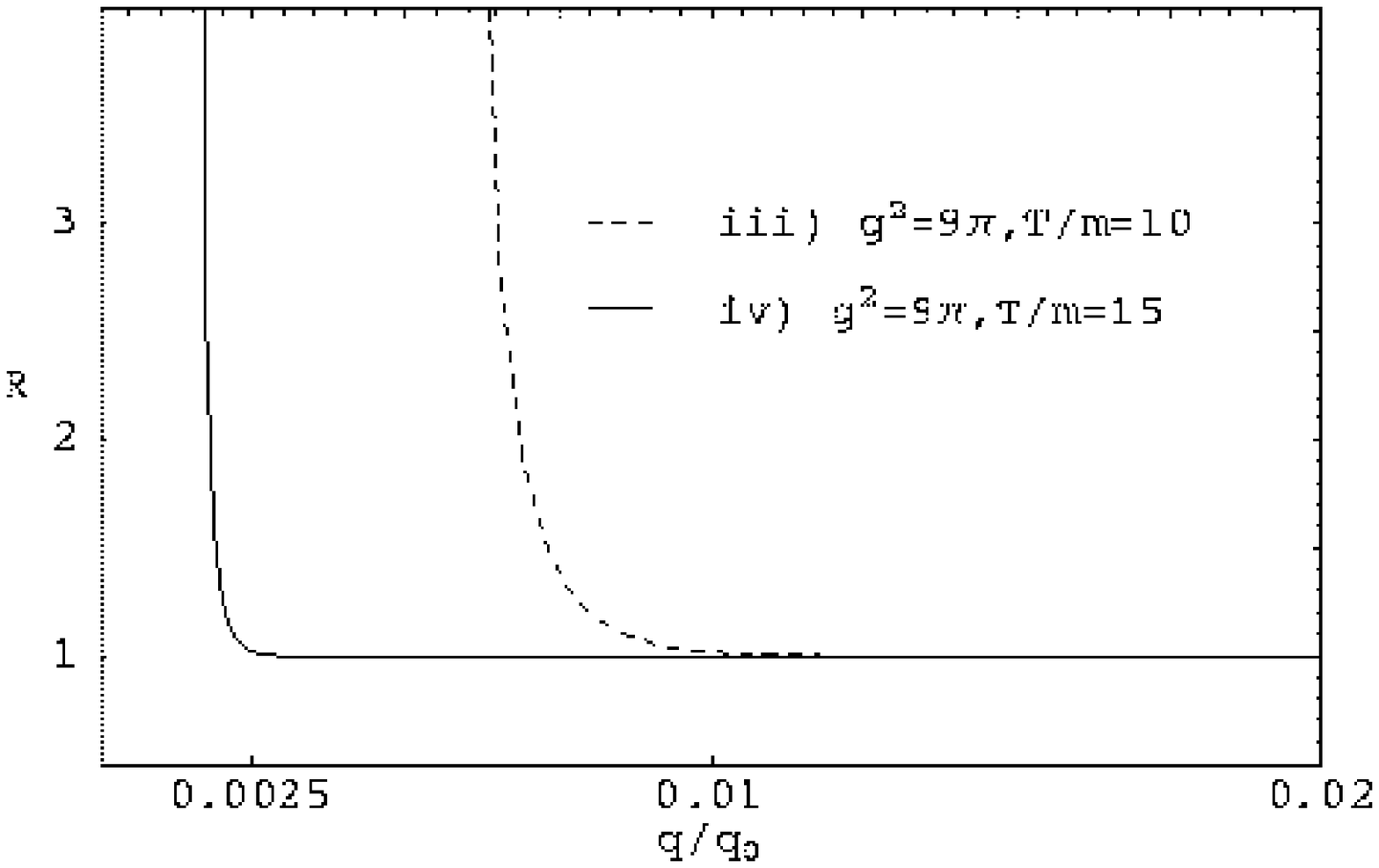,width=10cm,height=15cm}} \vspace*{-8.5cm}
\caption {\label{fig5}
 The two-point correlator ratio $R(T,g;q)$
  for different values of $T$ and $g^2$. The numerical parameters
  for the above curves are: i)$\hat z=0.225$, $(2\pi
  T)^{-1}/q_0=0.044$, ii) $\hat z= 0.082$, $(2\pi
  T)^{-1}/q_0\simeq 1.3 \times 10^{-4}$, iii) $\hat z=0.01$, $(2\pi
  T)^{-1}/q_0=2.3\times 10^{-5}$, iv) $\hat z=0.005$, $(2\pi
  T)^{-1}/q_0=5\times 10^{-6}$.}
\end{figure}

After conversion to $g$, our results for $R(g,T;X)$ are plotted in
Figure 
\ref{fig5} for
different values of $T$ and $g^2$. In all those curves, we have
considered values $q\gg (2\pi T)^{-1}$, so that we can neglect the
$\tau$  dependence and thus (\ref{r}) becomes just the ratio
$f_{+-}(q)/f_{++}(q)$. We also 
recall that the $f_{++}$ function can take negative values for smaller
$q$, but we have only displayed in Figure \ref{fig5} the region where
 $R(T,g;q)$ is  positive.

Let us  define the parameter $q_0$ to be the mean inter-particle
distance so that $q_0 = 1/f_+$, i.e.\ the inverse of the density
of charges, regardless of whether they are positive or negative. Thus,
$$
q_0(\lambda,T)=\frac{2T}{m\ll\overline\psi\psi\gg}
$$
with $m\ll\overline\psi\psi\gg$ in (\ref{cond}).

As in our previous analysis of the pressure and the
fermion condensate, the relevant parameter here is $\hat z$ in
(\ref{hatz}).
For very small $\hat z$ the system is in the condensed regime. Since we
need to work at large $g^2/\pi$ for DR to work,
small $\hat z$ means large $T$. In that regime, the $+-$ correlator is
much larger than the $++$ one at short distances (but yet large distances
compared to $(2\pi T)^{-1}$), as a result of the tendency of the
system to form $\sigma_+\sigma_-$ pairs (``molecules'').
 For large distances, $R$
tends to 1. In the ``normal'' phase, i.e, when no ``molecules'' are
formed, we expect that $R$ starts approaching its asymptotic value
near $q=q_0$ and it does not grow very much below $q_0$. Conversely
in the ``molecule'' phase, $R$ grows for small $q$ and it
approaches 1 much faster, defining a
certain $q_{cond}\ll q_0$, which we interpret as the screening length or
``molecule'' size. For the curves plotted in Fig.\ref{fig5}, we get 
 i) $q_{cond}\simeq 1.3 q_0$, ii)  $q_{cond}\simeq 0.3 q_0$,
iii) $q_{cond}\simeq 0.01 q_0$, iv) $q_{cond}\simeq 0.0025 q_0$. 
 One can clearly observe 
that  the screening length decreases with the temperature.

\section{Dimensional reduction of the Thirring model at
nonzero fermion chemical potential}\label{sec:chempot}

In \cite{gs98}, the issue of thermal bosonisation in the MT/SG
system was also studied for $\mu\neq 0$, where $\mu$ is the fermion
chemical potential associated with the conservation of the fermion
number density $\bar\psi\gamma_0\psi$ in the Thirring model (the
conserved quantity is the number of fermions minus antifermions).  It
has been shown that the SG model Lagrangian acquires an extra
topological term, interpreted as  $\mu$ times the number of
kinks minus antikinks. The purpose of this section is to show that in the
 DR  regime, the partition function of the Thirring
model at $T>0$ and $\mu\neq 0$ can also be related with the 1D Coulomb
gas.

We start by recalling the result (in perturbation theory in $m$)
obtained in \cite{gs98} for the partition function of the MT model 
 at $T>0$ and $\mu\neq 0$:
$$
Z_{MT}(T,\mu;L) = Z_{m=0} (T,\mu;L)Z_C(T,\mu;L)
$$
where $Z_{m=0} (T,\mu;L)$ is the massless Thirring  result derived in
\cite{AlvGom98} and is given by
$$
Z_{m=0} (T,\mu;L) = Z_0^F (T,L)\exp\left[\beta L
  \frac{\mu^2}{2(\pi+g^2)}\right]
$$
and
\be
Z_C(T,\mu;L) = \sumn \left(\frac{1}{n!}\right)^2
\left[\frac{\rho m}{2}\left(\frac{T}
{\rho}\right)^{\lambda^2/4\pi}\right]^{2n} F_{2n}(\lambda,T,\mu;L).
\label{pfthmu}
\ee
Here
\be
 F_{2n}(\lambda,T,\mu;L) = \prodjint
\prod_{k<j}\left[ Q^2(x_j-x_k)\right]^{\epsilon_j\epsilon_k
\lambda^2/4\pi} \exp\left[i\mu\frac{\lambda^2}{4\pi}\sum_{j=1}^{2n}
\epsilon_j q_j\right]
\label{Fdefmu}
\ee
and, as in (\ref{Fdef}), $\epsilon_j=+1$ for $j\leq n$,  $\epsilon_j=-1$
for $n<j\leq 2n$. 
Notice that the above
partition function is real, as it should, because of the symmetry
of the integrand under the simultaneous relabelling
$q_{1}\leftrightarrow q_{n+1}$, $\dots$, $q_{n}\leftrightarrow
q_{2n}$.

Once more, we need to analyse the regime in which we can dimensionally
reduce the $Q$ variables inside the above integral through
(\ref{qasym}). It turns out that the analysis in this case is much
simpler than in the case of the two-point correlators and indeed the
conclusion is that DR works in the same regime as
for the $\mu=0$ pressure, that is, (\ref{drworks}). We quickly sketch
the argument here: first, for small $\lambda$ one readily notes that
the leading order is $\mu$-independent since $\sum_j
\epsilon_j=0$. Second, for large $T$, one arrives to an expression
identical  to (\ref{pexpla}), with $f_2(\lambda)$ in (\ref{f2lam})
replaced now by
$$
 f_2 (\lambda,\mu)=\int_0^\infty dq\int_0^1 d\tau
\left[Q^2(\beta q,\beta\tau)\right]^{-\lambda^2/4\pi}
\cos \left[\frac{\beta\mu\lambda^2 q}{2\pi}\right]
%\label{f2lamu}
%\ee
$$
whose numerical analysis shows a similar behaviour than
$f_2(\lambda)$, i.e, its corresponding $\Delta f_2 (\lambda,\mu)$
  remains of $\Od(1)$ for $\lambda^2/4\pi\lsim 1$ for
different values of $\beta\mu$. That is easily understandable by
 realising that $f_2(\lambda,\mu)\leq f_2(\lambda)$ and
$f_2^{asym}(\lambda,\mu)>0$, so that $\Delta f_2 (\lambda,\mu)\leq
f_2(\lambda) $, which near $\lambda^2\lsim 4\pi$ (where the error is
bigger) is $f(\lambda^2\lsim 4\pi)=\Od (1)$. 

Given therefore that one can work in DR also for
$\mu\neq 0$, the $Q$'s in (\ref{Fdefmu}) may be replaced by (\ref{qasym}).
We then realise that $Z_C(T,\mu)$ in (\ref{pfthmu}) is now the
grand canonical classical partition function of the Coulomb gas,
the extra $\mu$-dependent term having exactly the form of a purely
imaginary external electric field term added to the Hamiltonian of the
system. In fact, note that if we  wanted to describe ($N$
positive and $N$ negative) charges on the
line  not only interacting among themselves with a Coulomb force, but
also with an external electric field ${\cal E}$, we should add to
 the Coulomb Hamiltonian  the term
\be
{\cal H}\rightarrow {\cal H}-\sigma{\cal E}\sum_{i=1}^{2N}\epsilon_i q_i.
\label{elect}
\ee
However, this is precisely the form of the $\mu$-dependent
term in (\ref{Fdefmu}), so that once the $Q$'s are  dimensionally
reduced one obtains
%\be
$$
Z_C(T,\mu;L)=\Omega
\left(z=\frac{m^2}{2T}\left(\frac{2T}{m}\right)^{\lambda^2/4\pi},
\theta=T,\sigma=\frac{\lambda T}{\sqrt{4\pi}},
{\cal E}=\frac{i\mu\lambda}{\sqrt{\pi}};L\right)
%\label{pfequivmu}
%\ee
$$
with $\Omega(z,\theta,\sigma,{\cal E},L)$ the grand canonical
partition function of the classical Coulomb gas interacting with an
external electric field ${\cal E}$ through (\ref{elect}).

The exact calculation of the pressure $P(z,\theta,\sigma,{\cal
  E})=\lim_{L\rightarrow\infty} \log\Omega/\beta L$ was also performed in
\cite{lenard1}.\footnote{Note
that the ensemble used in \cite{lenard1} was the canonical one at fixed
$N$ and $P$ (so that $L$ is varying, for instance by means of  a moving
piston)
 rather than the grand canonical ensemble at fixed $z$ and $L$ being
used here---the former are the conjugate variables of the latter.
However, in the thermodynamic limit the results
obtained in the different ensembles should be equivalent,  after the
standard identifications. Namely, for $N\rightarrow\infty$,
$L\rightarrow\infty$ with $N/L$ fixed, we can
identify $N$ with $\bar N$ and $\bar L$ with $L$,
where the bar denotes averages in the corresponding ensemble.}
In this case though, the result cannot be expressed in a simple way 
in terms of Mathieu functions as for $\mu=0$.  Now it is given by
\be
P_C(T,\mu)\equiv\lim_{L\rightarrow\infty}\frac{1}{\beta L}\log
Z_C(T,\mu;L)=\frac{2\pi T^2}{1 + g^{2}/\pi}\gamma(\hat z)
\label{pmu}
\ee
where $\hat z$ is given in (\ref{hatz}) and the function $\gamma(\hat
z)$ is defined implicitly as follows. Let $G(z,\gamma,\eta)$ be
the following continued fraction:
\be
G (z,\gamma,\eta)=\frac{z}{\gamma +2\eta
  +1-\frac{z}{\gamma+4\eta+4-\frac{z}{\gamma+6\eta+9-\dots}}}
\label{contfrac}
\ee
where 
$$\eta=\frac{{\cal E}}{4\pi\sigma}=i\frac{\mu}{2\pi T}, $$
and define
$$
\bar Q (z,\gamma,\eta)=\frac{1}{\gamma}\left[G (z,\gamma,\eta)+
G(z,\gamma,-\eta)\right].
$$
Now let $z_0(\gamma,\eta)$ be the smallest
positive solution to the equation
\be
\bar Q(z_0(\gamma,\eta),\gamma,\eta)=1.
\label{impli}
\ee
Then $\gamma (\hat z)$ is defined implicitly by  identifying 
$\hat z^2=z_0(\gamma,\eta)$.

First note that
for $\mu=0$ the above result reduces to the solution in terms of
Mathieu functions we have found in section \ref{sec:pf}. In that case
equation (\ref{impli}) defines precisely the condition that the two
parameters $\hat z$ and $\gamma$ of Mathieu's differential equation
(\ref{mathieu}) should satisfy in order that the solution is $2\pi$
periodic. Therefore for $\mu=0$, we have just $\gamma=\gamma_0$.
 In \cite{lenard1} it was shown that equation (\ref{impli}) has
 always a solution. It is worth mentioning that in our case, due to
 the purely imaginary character of $\eta$, the convergence
 of the different integrals analysed in \cite{lenard1} and hence of
 the continued fraction (\ref{contfrac}) is
 automatically ensured.  This should be contrasted with
the real $\eta$ case where $\gamma$
 and $\eta$ have to satisfy certain constraints to ensure
 convergence. In addition,
 notice that for $\gamma$ real, $\bar Q$ in (\ref{contfrac}) is
 also real, which means that the pressure is real, as it should.

Equation (\ref{impli}) can be solved exactly in the asymptotic limit
 $\hat z\ll 1$ and $\gamma\ll 1$, where we know  there is
 condensation into ``molecules''. For $z,\gamma\ll 1$,
 $\bar Q(z,\gamma,\eta)$ in (\ref{contfrac}) just becomes
$$
\bar Q(z,\gamma,\eta)\simeq \frac{2z}{\gamma(1-4\eta^2)}
$$
so that
\be
\hat z^2\simeq \frac{\gamma (\hat z)}{2}\left[1-4\eta^2\right]
\label{asrel}
\ee
is the smallest  solution to $\bar Q=1$ and defines $\gamma (\hat z)$
in this limit.
Therefore, using (\ref{pmu}),  replacing the values for $\hat z$
and $\eta$, and eliminating $\lambda^2$ for $g^2$, 
we find in the $\hat z\ll 1$ limit (condensation regime)
$$
P_C(T,\mu)\simeq \frac{m^4(1 + g^{2}/\pi)}{4\pi T^2}
\left( \frac{2T}{m}\right)^{2(1 + g^{2}/\pi)^{-1}}\frac{1}{1+\mu^2/\pi^2
  T^2}.
%\label{pmucond}
$$

One can check that for $\mu=0$ the above result reduces to the one we
have found in
section \ref{sec:pf} in that regime. Notice that the pressure not only
 vanishes for large $T$ ($T\gg m$) but also for large $\mu$ at
large $T$ ($\mu\gg T\gg m$). That is, chiral restoration also takes
place for large $\mu$ (in the large $T$ limit here). This is confirmed by
differentiating the above with respect to $m$ to get the fermion
condensate, which clearly vanishes as $\mu\rightarrow\infty$. On the
other hand, differentiating the pressure with respect to $\mu$ yields the net
averaged fermion density,
$$
\rho (T,\mu)=\lim_{L\rightarrow\infty}\frac{1}{L}\ll
\overline\psi\gamma^0\psi\gg \; =\frac{\partial}{\partial\mu} P(T,\mu).
$$
Thus, we get for the total fermion density in the condensation regime
\be
\rho (T,\mu)=\frac{\mu}{\pi + g^2}-\frac{m^4(1 + g^{2}/\pi)}{2 \pi^3 T^4}
\left( \frac{2T}{m}\right)^{2(1 + g^{2}/\pi)^{-1}}
\frac{\mu}{(1+\mu^2/\pi^2 T^2)^2}
\label{rhocond}
\ee
where the first contribution is the massless result obtained in
\cite{AlvGom98}.  The above
result for the fermion density deserves some comments. One notes
that  in (\ref{rhocond})
the Coulomb correction to the massless case is always negative and 
 very small in this regime, since the above result is valid only when $T\gg
m$. 
Therefore as $T\rightarrow\infty$, only the massless contribution
remains since the first term is
$T$-independent. This is
nothing but another signal of chiral restoration for large $T$: in
that regime there remain only massless fermion excitations in the
thermal bath.   
The same is true if we take in (\ref{rhocond})
the limit $\mu\gg T$ as commented above. In our picture of chiral
condensation, the Coulomb  term in (\ref{rhocond})
represents the density of ``residual'' charges that are not yet
paired to form chiral  ``molecules''.

For arbitrary values of $\hat z$, the solution to (\ref{impli}) has to
be found numerically. First, one truncates the continued fraction
(\ref{contfrac}) to a given order, making sure that the difference
between successive truncations becomes smaller than a fixed numerical
 precision (see \cite{abramo} for details).
This procedure is meaningful only if the continued
fraction is convergent and indeed, by computing it in this way we
check numerically the arguments about convergence 
discussed above.  Then one has to find, also numerically,  the
smallest positive zero  of $\bar Q(z_0,\gamma,\eta)=1$. 
For that purpose we have used a bisection method. 
In fact, since our objective is to get the
function $\gamma(\hat z)$, it is simpler to look for the zeroes in
$\gamma$ fixing $z_0=\hat z^2$. Besides, we have made use of the asymptotic
expression (\ref{asrel}) to check the answer in the large $\hat z$ limit.

\begin{figure}
\hspace*{-1cm}
\vspace*{-1.5cm}
\psfig{figure=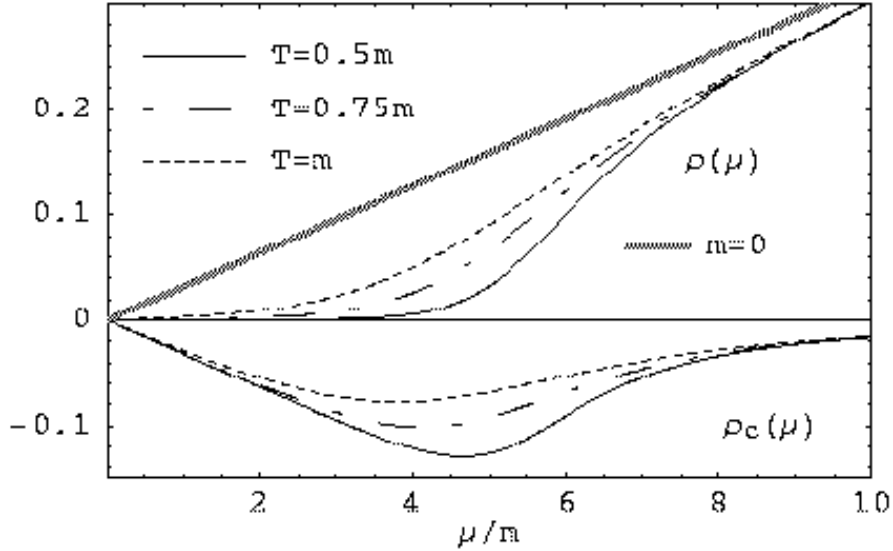,height=21.5cm}
\vspace*{-10cm}
\caption {\label{fig6}
The total net fermion density $\rho (\mu)$ (above the $\mu$ axis) and
the Coulomb gas correction  $\rho_C (\mu)$ to the massless gas (below
the axis) for $g^2/\pi=9$ and different  values of the
temperature around $T=m$. The massless result
$\rho_{m=0}(\mu)=\mu/(\pi+g^2)$ is also displayed for $g^2/\pi=9$.} 
\end{figure}

Our results for $\rho_C(\mu)$ and $\rho (\mu)$ are displayed in figure
\ref{fig6} for temperatures around $T\simeq m$
at fixed $g^2/\pi=9$ $(\lambda^2/4\pi=0.1)$.
When $\hat z$ increases, or equivalently when the temperature decreases,
 $\rho_C(\mu)$ (the curves below the axis in
 figure \ref{fig6}) has a similar shape as
that given by the second term in  (\ref{rhocond}), but  its magnitude
increases so that the (negative) 
correction to the massless gas becomes more and
more important. In fact,  we see that  for $\mu\lsim m$, $\rho_C(\mu)$ 
roughly equals the massless contribution, thus yielding
$\rho(\mu)\simeq 0$ in that region,
whereas   for large $\mu$, $\rho_C(\mu)$ becomes negligible
 and only the massless contribution remains. 

In view of this behaviour, let us define two ``critical'' points:
$\mu_0\simeq m$ 
such that $\rho(\mu\lsim \mu_0)\simeq 0$,  $\rho(\mu> \mu_0)\neq
0$, and $\mu_1$ such that $\rho(\mu>\mu_1)\simeq
\rho_{m=0}(\mu)$, where the curve for $\rho(\mu)$ catches up with 
the massless one, i.e.\  $\mu_1$ is the chiral restoration point.
  Recall that we have no true critical behaviour here,
so that $\mu_0$ and  $\mu_1$ can be defined only approximately.  

Physically, $\mu_0$ at  $T=0$ 
would represent  the energy needed to excite a fermion in the thermal
bath. Thus the behaviour of $\rho(\mu)$ in figure \ref{fig6} 
 near $\mu=\mu_0$ is typical of a
massive Fermi gas:  It is approximately zero until $\mu\simeq \mu_0$. 
For instance,  notice that for a free gas of
fermions of mass $m$, $\mu_0(T=0)=m$ and 
$\rho(\mu,T=0)=(\mu-m)\theta(\mu-m)$.  
 For a small but nonzero 
temperature, that characteristic
step-function behaviour is smoothed because there is also thermal
energy available, and this can also  be observed in figure \ref{fig6} as
$T$ increases. Remember that  $T\simeq m$ is the
lower temperatures we  can take  in our DR regime---see our previous 
comments about the problems with the $T\rightarrow
0^+$  limit within our approach. In addition, 
notice that in our case, the almost exact 
cancellation between $\rho_C(\mu\lsim\mu_0)$ and 
$\rho_{m=0}(\mu\lsim\mu_0)$ for low $T$ takes place after a nontrivial 
numerical calculation of the Coulomb part, and therefore it provides a
good consistency check. 

Let us now comment about the behaviour we observe for
$\mu>\mu_1$ in figure \ref{fig6}. 
 Following our previous description of  chiral
symmetry restoration in terms of pairing of condensates, our results
suggest that the chemical potential also increases the effective
``coupling charge'' holding the condensate pairs together.  
 In fact, from the viewpoint of the Coulomb gas, we now have another way
 of increasing the mean potential energy with respect to the mean
 kinetic energy; namely to switch on a
 strong  electric field $\vert{\cal E}\vert \gg T$. It seems natural
 then that in this regime the systems behaves as in the low
 temperature phase of Lenard, forming ``molecules''. This corresponds to
 the $\mu>\mu_1$ regime in figure \ref{fig6}, i.e.\ moderate
 temperatures $T\simeq m$ and large chemical potentials $\mu\gg T$.

As $T$
increases further, the effective mass of the fundamental fermion decreases
with $T$, until, for very large $T$, the chiral symmetry is
effectively restored and then $\rho(\mu)$  grows linearly with
$\mu$ $\forall\mu>0$ as given by (\ref{rhocond}).

 This type of 
behaviour also takes place
  in QCD at finite baryon density $\rho_B$ \cite{stephan}.  
There, at $T=0$ one has 
 $\mu_0\simeq m_B\simeq$ 930 MeV,  the vacuum to nuclear matter
transition point, with $\rho_B (\mu_0^+)\simeq$ 0.16 fm$^{-3}$ 
the density of nuclear matter, and at $\mu_1\simeq$ 1300 MeV (for
$T=0$) the
chiral symmetry is restored (vanishing quark condensate). At both points 
it is expected that the phase transition is of first order at $T=0$.

 Another system where one
 observes a similar behaviour of $\rho (\mu)$ is the massive
Schwinger model \cite{fikosu79},
where for large $\mu$ the system is in the deconfined phase, unlike
the massless case,  where there is confinement for any 
$\mu$ \cite{AlvGom98}.

\section{Conclusions}

In this paper we have considered the massive Thirring model in 1+1
space-time dimensions at finite temperature $T$ and chemical potential
$\mu$. We have shown that in a certain regime, which we have 
denoted dimensional
reduction, the statistical mechanics of this system is the same as that
of a classical Coulomb gas in one spatial dimension, where the unit
charge grows linearly with the temperature.

The range of validity of the DR regime depends on the observable under
consideration: for the pressure, the fermion density and the fermion
condensate, we have seen that it works both for
 $T\gg m$ with $g^2>0$ and for $T\gsim m$ with
 $g^2/\pi\gg 1$ (strong coupling regime).
However, for the two-point correlators of
 fermion chiral operators $\sigma_\pm (x)$, it only works for large $g$,
 requiring in addition that the spatial distance $q$ between the two
 chiral operators is such that $q\gg \beta/2\pi$.

Thanks to the analogy with the Coulomb gas, we have been able to
calculate exactly the pressure, fermion density, fermion condensate and
two-point correlators of the MT model. Our results show that the
chiral symmetry is restored both for high $T$ and high $\mu$ in a
continuous way (no phase transition). Chiral symmetry restoration
takes place only in the strong coupling (large $g^2$) regime. 
The symmetry restored phase corresponds to the Coulomb ``condensed''
phase, in which $+-$ pairs of charges tend to pair, forming
``molecules'', whereas our low $T$ phase is the Coulomb ``plasma''
phase, in which the charges are free. The low $T$ and high $T$
behaviours are reversed in our case with respect to the Coulomb gas
because  the coupling between charges grows with $T$. 
We have also seen
that the r\^ole of the
Coulomb charges is played here by the  $\sigma_\pm (x)$
operators. These operators tend to pair in the high $T$ or  $\mu$
phase forming chiral invariant $\sigma_+\sigma_-$ combinations.  These
play the r\^ole of the ``molecules'', whose ``size'' is determined by
the condensation or screening length, which  we have estimated for
different values of the coupling constant and temperature. We find
that the screening length decreases with $T$. 
In the restored phase, this
screening length is of the same order of the inter-particle distance,
which is proportional to   the inverse of the fermion condensate.

The case of a nonzero fermion chemical potential $\mu$ can also be
related to the 1D Coulomb gas, by noting that $\mu$ plays the r\^ole of a
purely imaginary external electric field. For high $T$, the system
behaves with $\mu$ as the massless case, which is consistent again
with chiral restoration since the effective mass falls off with $T$. For
$T\simeq m$, we see that the fermion density is very small until
$\mu_0\simeq m$ and then starts growing (typical massive Fermi gas
behaviour) until eventually, for large
$\mu$ ($\mu>\mu_1$), it reaches again the massless linear behaviour (chiral
restoration). This  is again consistent with the idea that
the fermion excitation of the system becomes massless  both at high
$T$ and  at high $\mu$. In the case of moderate temperatures, chiral
condensates are held together at large $\mu$ by means of the strong
external electric field.  

Among the analogies with QCD we have found, it is worth mentioning
that the chiral condensate vanishes  for both large $T$ and $\mu$,
although in our case there is no phase transition. In addition, 
 the type  of behaviour of the fermion density 
$\rho (\mu)$ we have obtained  at low temperatures is
 also quite similar to that of the baryon number density $\rho_B$ 
 in QCD, where $\mu_0$ and $\mu_1$ are the vacuum to
 nuclear matter and chiral restoration critical points, respectively.

\section*{Acknowledgments}

We would like to thank Tim Evans for numerous helpful discussions and
suggestions. 
A.G.N has received
support through CICYT, Spain, project AEN97-1693.
D.A.S.\ is supported by P.P.A.R.C.\ of the UK through a
research fellowship and is a member of Girton College, Cambridge.
This work was supported in part by the E.S.F.

\appendix
\section{Dimensional reduction for the two-point
correlators}

In order to
 compare the  $\sigma\sigma$ correlators
(\ref{sigma+-sg})-(\ref{sigma++sg}) with the corresponding
$ff$ ones ((\ref{f+-lenard})-(\ref{f++lenard})) in the Coulomb gas,
the former must be dimensionally reduced through (\ref{qasym}) just as
in the case of the partition function.
All calculations are performed with respect to $\lambda^{2}$.

Let us start with the $+-$ correlator.
The relevant integral to analyse now is:
$$
 F_{2n}^{+-}(\lambda,T,L;X)=\prodjint
\prod_{k<j}\left[
  Q^2(x_j-x_k)\right]^{\epsilon_j\epsilon_k\lambda^2/4\pi}
\left[\frac{Q^2(x_j-X)}{Q^2(x_j)}\right]^{\epsilon_j \lambda^2/4\pi}
$$
where $X\equiv (q,\tau)$, 
$0\leq q\leq L$ and we have set $Y=0$. The above integral is the
counterpart of (\ref{Fdef}). Let us  consider first the
limit $\lambda^2/4\pi\ll 1$. We readily realise that
the $X$-dependent  factor in the integrand does not contribute to
leading order, since $\sum \epsilon_j =0$. Thus, following similar
steps as in section \ref{ssec:smallla}, we get
$$
\frac{\ll T_C \sigma_+ (X)\sigma_- (0)\gg-
\ll T_C \sigma_+ (X)\sigma_- (0)\gg^{asym}}
{\ll T_C \sigma_+ (X)\sigma_-
  (0)\gg}=-\frac{\lambda^2}{4\pi}\frac{m^2}{2T^2}
\lim_{\eta\rightarrow\infty}\eta \hat I (\eta)+
\Od\left(\frac{\lambda^2}{4\pi}\right)^2
$$
where $\hat I(\eta)$ is given in (\ref{ieta}) and
 the superscript $asym$ means now replacing the $Q$'s inside the
integrals as (\ref{qasym}), but {\it not} for the function $Q^2(X)$
outside, which would be allowed only for 
$q \gg \beta/2\pi$ (see our comments in the main text). Notice
 once more, that one should be careful when taking $T$ to arbitrary small
 values at the same time as expanding in $\lambda$. Here, we even find
 a term $m^2/T^2$ to leading order. Again, we will
 consider our approach valid only for $T\gsim m$.

Our numerical analysis of $\hat I (\eta)$ shows that it vanishes much
faster than $\eta^{-1}$ as $\eta\rightarrow\infty$, so that we see
that for small $\lambda$, DR (for the integrals) is
justified.
Notice that the above relative error is scale independent, so we have
fixed $\rho=m$. That is not true for the correlator (\ref{sigma+-sg})
itself, where there is an explicit $\rho^{2-\lambda^2/2\pi}$ overall
 dependence.

Next, we will discuss the $T\gg m$ case. Following the same steps as
for the partition function, we get now
\ba
\lefteqn{
\ll T_C \sigma_+ (X)\sigma_- (0)\gg=\frac{Z_0^F(T)}{Z_{MT}(T)}
\left(\frac{T}{\rho}\right)^{\lambda^2/2\pi}
\left(\frac{\rho}{2}\right)^2\left[Q^2(X)\right]^{-\lambda^2/4\pi}
}&&
\nonumber\\
&\times&
\left\{1+\left[\frac{1}{2}\left(\frac{m^2}{T^2}\right)^{1-\lambda^2/8\pi}
\right]^2 f_2^{+-}
(\lambda,X)+\Od\left(\frac{m}{T}\right)^{8-\lambda^2/\pi}
\right\}
\label{highTexp+-}
\ea
with
\be
f_2^{+-}(\lambda,X)=\int_0^1 d\tau_1 d\tau_2 \int_0^\eta dq_1 dq_2
\left[\frac{Q^2(q_1-q,\tau_1-\tau)Q^2(q_2,\tau_2)}
{Q^2(q_1-q_2,\tau_1-\tau_2)Q^2(q_2-q,\tau_2-\tau)Q^2(q_1,\tau_1)}
\right]^{\lambda^2/4\pi}
\label{i2+-}
\ee
where $X=(q,\tau)$, $q$ has been rescaled $q\rightarrow q/\beta$ and
all the $Q^2$ functions  are evaluated at $\beta=1$.  Notice
that we have replaced $\rho=m$ in (\ref{highTexp+-}) only for the
scale-independent part, keeping the scale dependence outside the curly
brackets.

Thus, for $T\gg m$, we have to analyse the behaviour of the integral
(\ref{i2+-}) with $\lambda$ and $X$ or, rather, the difference
between $f_2^{+-}(\lambda,X)$ and $f_2^{+-,asym}(\lambda,X)$ obtained by
replacing the $Q$'s by their asymptotic values. Notice that
$f_2^{+-}(\lambda,0)$ gives exactly the same contribution as the partition
function evaluated to the same order, that is,
$f_2^{+-}(\lambda,0)=f_2(\lambda)$ in (\ref{f2lam}).  
In fact, in the language of
Feynman diagrams, this is a disconnected
contribution, proportional to $\eta$, so that the factor $Z_0/Z_{MT}$
in front ensures that the answer for this correlator is finite as
$\eta\rightarrow\infty$ and we take it into account by subtracting the
$X=0$ contribution to the correlators. The same is true   for
$f_2^{+-,asym}(\lambda,X)$.   Bearing this in mind, let us concentrate on
the values of $\lambda^2$ close to and below $4\pi$. The reason is
twofold: On the one
hand, we have seen already that for small $\lambda$ the DR works. On
the other hand, those are the
values that will give us the biggest contributions to the error, since
the integrand of the asymptotic form is not singular at any point,
whereas in (\ref{i2+-}), there are singular regions where the
denominator vanish, and those give larger (but finite) contributions to the
integral the closer we approach to $\lambda^2\lsim 4\pi$.
 The direct numerical evaluation of (\ref{i2+-}) is a hard task.
However, for $T\gg m$, all we need to show is that the
 difference with its asymptotic value remains
 bounded close to  $\lambda^2\lsim 4\pi$. Therefore, let us estimate
 it by looking only at the biggest contributions, i.e, those of the
 regions close to the singularities of the integrand. Firstly, in the
 region $x_1\simeq x_2$, where $x_1=(q_1,\tau_1)$ and
 $x_2=(q_2,\tau_2)$,
 we see that the integrand simply tends to its
 $X=0$ value, and hence that contribution cancels with the partition
 function. On the other hand, near $x_2\simeq X$ or $x_1\simeq 0$, the
 integrand goes like $\left[Q^2(X)/
 Q^2(x_2-X)Q^2(x)\right]^{\lambda^2/4\pi}$. Then,
\be
f_2^{+-}(\lambda,X)-f_2^{+-}(\lambda,0)
\sim \left[Q^2(X)\right]^{\lambda^2/4\pi}
\left[f_2(\lambda)\right]^2\qquad \mbox{(dominant contribution)}
\label{dom+-}
\ee

 We have seen in
section \ref{ssec:hight} that
$\vert f_2(\lambda)-2^{\lambda^2/2\pi}/2\lambda^2\vert \lsim 1$ for
$\lambda^2<4\pi$.

As for the asymptotic contribution to the integral (\ref{i2+-}), we
can evaluate it explicitly by integrating over the four separate
regions on the $(q_1,q_2)$ plane according to the different
relative signs of $q_1-q_2$, $q_1-q$ and $q_2-q$. We obtain in the
$\eta\rightarrow\infty$ limit
\be
f_2^{+-,asym}(\lambda,X)-f_2^{+-,asym}(\lambda,0)=
\frac{2^{\lambda^2/2\pi+1}}{\lambda^2}
\left\{-\frac{8}{3}q+\frac{2}{\lambda^2}\left[e^{\lambda^2q/2}+
\frac{1}{9}e^{-3\lambda^2q/2}-\frac{10}{9}\right]\right\}
\label{as+-}
\ee
Therefore, from (\ref{as+-}), (\ref{dom+-}) and (\ref{highTexp+-}), we
see that the relative error for the two-point $+-$ correlator---which
is a scale independent quantity---does {\it not}
remain bounded at large $T$ for $\lambda^2$ close to $4\pi$, unlike
the pressure, but it grows arbitrarily large for $q\gg 1$.
 In particular, taking
 $\lambda^2\lsim 4\pi$ and $q\gg 1$ in the above
 expressions, we find
$$
\frac{\ll T_C \sigma_+ (X)\sigma_- (0)\gg-
\ll T_C \sigma_+ (X)\sigma_- (0)\gg^{asym}}
{\ll T_C \sigma_+ (X)\sigma_-
  (0)\gg}\simeq c\frac{m^2}{T^2}\exp [2\pi q]+
\Od\left(\frac{m^2}{T^2}\right)^2
$$
with
$c=(f_2(\lambda)^2/4-1/\pi^2)/4$.

Now we turn to  the $++$ correlator. As before, we shall start with the limit
$\lambda^2/4\pi\ll 1$. From (\ref{sigma++sg}) we have now
\ba
\ll T_C \sigma_+ (X)\sigma_+ (Y)\gg&=&
\frac{Z_0^F(T)}{Z_{MT}(T)}\left(\frac{T}{\rho}\right)^{\lambda^2/2\pi}
\left(\frac{\rho}{2}\right)^2 \sumn \frac{1}{n!(n+2)!}
\left[\frac{m^2}{2}\left(\frac{T}
{m}\right)^{\lambda^2/4\pi}\right]^{2n+2} \nonumber\\
&\times&
\left[Q^2(q,\tau)\right]^{\lambda^2/4\pi}F_{2n}^{++}(\lambda,T,X)
\label{++la}
\ea
with
$$
F_{2n}^{++}(\lambda,T,X)=\beta^{2(2n+2)}\prod_{j=1}^{2n+2}\int_0^1
d\tau_j\int_0^\eta
dq_j\left[Q^2(x_j-X)Q^2(x_j)\right]^{\epsilon_j\lambda^2/4\pi}
\prod_{k<j}\left[
  Q^2(x_j-x_k)\right]^{\epsilon_j\epsilon_k\lambda^2/4\pi}
$$
where now
$$
\epsilon_j=\left\{\begin{array}{ll}+&j=1,\dots,n\\-&j=n+1,\dots,2n+2
\end{array}\right.
$$
and, as before, we have set  $q\rightarrow q/\beta$, $\rho=m$, 
 in the scale-independent part, 
$X=(q,\tau)$ with $0\leq q\leq\eta$ and  the $Q$'s in the
integrand are evaluated at $\beta=1$. Notice that now
$\sum_j \epsilon_j\neq 0$, so that we find
an additional contribution when expanding in $\lambda^2/4\pi$. We find
for the relative error
\ba
\frac{\ll T_C \sigma_+ (X)\sigma_+ (0)\gg-
\ll T_C \sigma_+ (X)\sigma_+ (0)\gg^{asym}}
{\ll T_C \sigma_+ (X)\sigma_+
  (0)\gg}=\frac{\lambda^2}{4\pi}\lim_{\eta\rightarrow\infty}
\left[\frac{1}{2}\hat I(\eta)-2\hat I(\eta,X)\right.
\nonumber\\
-\left.\frac{g(z)}{f(z)}\hat
  I(\eta)\right]+
\Od\left(\frac{\lambda^2}{4\pi}\right)^2
\nonumber
\ea
 with $z=m^2\eta/T^2$,
$$
\hat I(\eta,X)=\frac{1}{\eta}\left[\int_0^\eta dq_1\int_0^1 d\tau_1
\log\left[Q^2(q_1-q,\tau_1-\tau)\right]+2\eta \log 2-\pi\eta^2+2\pi
q(\eta-q)\right]
$$
and
\ba
f(z)&=&\frac{z}{2}\frac{d}{dz}\left[\frac{2}{z}I_0'(z)\right]\nonumber\\
g(z)&=&\left(\frac{z}{2}\right)^3\frac{d}{dz}\left[\left(\frac{2}{z}\right)^2
  f(z)\right]\nonumber
\ea
In the thermodynamic limit $\eta\rightarrow\infty$, it is easy to
check that  $\hat I(\eta,X)\rightarrow
\hat I(\eta)$ because of translation invariance, and that
$g(z)/f(z)\rightarrow z/2$, so that the relative error for small
$\lambda$ tends to zero (for $T\gsim m$). Again, $asym$  means the same as for
the $+-$ correlator, i.e, replacing the $Q$'s by their asymptotic
values only inside the integrals, but not for the
$\left[Q^2(q,\tau)\right]^{\lambda^2/4\pi}$ factor in (\ref{++la}).

Finally, we will consider $T\gg m$ for this correlator. We have now
\ba
\ll T_C \sigma_+ (X)\sigma_+ (0)\gg&=&\frac{Z_0^F(T)}{Z_{MT}(T)}
\left(\frac{T}{\rho}\right)^{\lambda^2/2\pi}
\left(\frac{\rho}{2}\right)^2\left[Q^2(X)\right]^{\lambda^2/4\pi}\nonumber\\
&\times&\left[\frac{1}{2}\left(\frac{m^2}{T^2}\right)^{1-\lambda^2/8\pi}
\right]^2 f_2^{++}(\lambda,X)
\Od\left(\frac{m}{T}\right)^{8-\lambda^2/\pi}
%\label{highTexp++}
\nonumber
\ea
with
\ba
f_2^{++}(\lambda,X)&=&\int_0^1 d\tau_1 d\tau_2 \int_0^\eta dq_1 dq_2
\nonumber\\&\times&
\left[\frac{Q^2(q_1-q_2,\tau_1-\tau_2)}
{Q^2(q_1-q,\tau_1-\tau)Q^2(q_2-q,\tau_2-\tau)Q^2(q_1,\tau_1)Q^2(q_2,\tau_2)}
\right]^{\lambda^2/4\pi}
%\label{i2++}
\nonumber
\ea
Notice that
there is an important difference between this correlator and the
previous cases we have analysed, namely, that the difference between
the actual value and the asymptotic limit shows up already to leading
order in the $m/T$ expansion, which is now
$\Od(m/T)^{4-\lambda^2/2\pi}$ rather than $\Od(1)$.
 That means that the relative error for the correlator
 $f_2^{++}(\lambda,X)$
starts now at  $\Od(1)$, and therefore is not bounded.  To check
that this is indeed the case,
it is enough  to take
the dominant contribution to the integral by picking up the different
poles,  as we did before with the $+-$ correlator.
 One easily finds now
$$
f_2^{++}(\lambda,X)\sim
\left[Q^2(X)\right]^{-\lambda^2/4\pi}
\left[f_2(\lambda)\right]^2\qquad \mbox{(dominant contribution)}
%\label{dom++}
$$

Notice that in this correlator there is no disconnected contribution,
which is consistent with the fact that its large $T$ expansion begins
at NLO whereas that of the partition function starts at $\Od(1)$. In
other words, $f_2^{++}(\lambda,X)$ is finite as
$\eta\rightarrow\infty$. On the other hand, the asymptotic
contribution yields
$$
f_2^{++,asym}(\lambda,X)=-\frac{2}{\lambda^2}\left[2qe^{-\lambda^2
    q}+3\frac{e^{-\lambda^2 q}}{\lambda^2}-4\frac{e^{-\lambda^2
      q/2}}{\lambda^2}\right]
$$
Therefore, one readily checks that the relative error for  
$f_2^{++} (\lambda,X)$ is
indeed $\Od (1)$ for arbitrary  $q$, so that we cannot bound it
 in the $T\gg m$ limit.

Summarising, the structure of the correlators
(\ref{sigma+-sg})-(\ref{sigma++sg}) makes it possible to reduce them
dimensionally in the small $\lambda$ limit but not in the large $T$
one. As commented above, DR will also work for the
factors $Q^2(X)$ outside the integral in the limit
$q \gg \beta/2\pi$.

%\newpage
\typeout{--- No new page for bibliography ---}

\end{document}